\documentclass[prd,aps,nofootinbib,floatfix,11pt]{revtex4}
\usepackage{amsmath,graphicx,epsfig,amssymb,dsfont,mathtools,}
\usepackage[usenames]{color}
\usepackage{ulem} 
\usepackage{bigstrut}
\usepackage{slashed}
\usepackage{multirow}
\usepackage{subfigure}

\allowdisplaybreaks

\begin{document}
\title{Heavy baryon decays into light meson and dark baryon within LCSR }

\author{Yu-Ji Shi$^{1,2}$~\footnote{Email:shiyuji@ecust.edu.cn (corresponding author)},
  Ye Xing~$^{3}$~\footnote{Email:xingye\_guang@cumt.edu.cn}
  and Zhi-Peng Xing~$^{4}$~\footnote{Email:zpxing@sjtu.edu.cn}}
\affiliation{$^1$ School of Physics, East China University of Science and Technology, Shanghai 200237, China\\
$^{2}$ Shanghai Key Laboratory of Particle Physics and Cosmology, \\
School of Physics and Astronomy, Shanghai Jiao Tong University, Shanghai 200240, China\\
$^3$ School of Materials Science and Physics, China University of Mining and Technology, Xuzhou 221116, China\\
$^{4}$ Department of Physics and Institute of Theoretical Physics, Nanjing Normal University, Nanjing, Jiangsu 210023, China}

\begin{abstract}
We studied the decays of Heavy baryon into a pseudoscalar meson and a dark baryon in the recently developed $B$-Mesogenesis scenario, where the two types of effective Lagrangians proposed by the scenario are both considered. The decay amplitudes of $\Lambda_b^0$ are calculated by light-cone sum rules using its light-cone distribution amplitudes. The decay amplitudes of $\Xi_b^{0,\pm}$ are related with those of $\Lambda_b^0$ through a flavor SU(3) analysis. The uncertainties of threshold parameter and the Borel parameter are both considered in the numerical calculation. The values of effective coupling constants in the $B$-Mesogenesis are taken as their upper limits that obtained from our previous study on the inclusive decay. The upper limits of the decay branching fractions are presented as functions of the dark baryon mass.
\end{abstract}
\maketitle

\section{Introduction}
The Standard Model of particle physics and cosmology have been proven to be successful in  describing the physics of the most microscopic and macroscopic worlds. However, there is inconsistency between these two models, where two of the most confusing problems are the existence of dark matter (DM) and the asymmetry of matter and anti-matter. Nowadays, a number of mechanisms aiming to solve this puzzle have been proposed according to
the Sakharov conditions for the baryogenesis \cite{Sakharov}. One of the disadvantages among these mechanisms is the existence of high energy scales and extremely massive particles, which make them difficult to be tested in experiments. Recently, a new mechanism called $B$-Mesogenesis is proposed by Refs.~\cite{Elor:2018twp,Alonso-Alvarez:2021qfd,Elahi:2021jia}, which aims to explain both the relic dark matter abundance and the baryon asymmetry without introducing high energy scales. The $B$-Mesogenesis is testable at hadron colliders and $B$-factories \cite{Alonso-Alvarez:2021qfd,Borsato:2021aum}, and also indirectly testable at the Kaon and Hyperon factories \cite{Alonso-Alvarez:2021oaj,Goudzovski:2022vbt}. Recently, the  Belle-II collaboration and the LHCb collaboration have started to search for the decays of $B$ mesons with energy missing according to the $B$-Mesogenesis \cite{Belle:2021gmc,Rodriguez:2021urv}.

The $B$-Mesogenesis proposes a new mechanism for Baryogenesis and DM production. In this model, during a late era in the history of the early universe, a certain heavy scalar particle $\Phi$ decays into $b,\bar b$ quarks, which then  form charged and neutral $B$-mesons after the universe cool down. After that, the neutral mesons $B^0, {\bar B}^0$ quickly undergo CP violating oscillations, and the remained mesons continue to decay into a dark sector baryon $\psi$ with baryon number $-1$ and visible hadron states with baryon number $+1$.  As a result, the asymmetry of the baryon and anti-baryon number from the CP violation during $B^0-{\bar B}^0$ oscillations is induced but without violating the total baryon number. 
Recently, there are a number of theoretical studies on the $B$ meson decays in the $B$-Mesogenesis. The exclusive decay $B\to p \psi$ was firstly studied by Ref.\cite{Khodjamirian:2022vta} using leading twist light-cone sum rules (LCSR) calculation, and a higher twist contribution are calculated in Ref.\cite{Boushmelev:2023huu}. A more complete study on $B$ meson decays into an octet baryon or  a charmed anti-triplet baryon plus $\psi$ was given in Ref.\cite{Elor:2022jxy}. In addition, a similar exclusive decay of $B$ meson into a baryon plus missing energy are studied by Ref.\cite{Dib:2022ppx} for probing the lightest neutralino. Besides the exclusive decays, previously we studied the semi-inclusive decay of $B\to X_{u/c,d/s} \psi$ using heavy quark expansion (HQE), where $X_{u/c,d/s}$ denotes any possible hadron states containing $u/c$ and $d/s$ quarks with unit baryon number \cite{Shi:2023riy}. Using  the experimental  upper limits on the branching fractions of $B\to X_{u/c,d/s} \psi$ from the ALEPH experiment \cite{ALEPH:2000vvi,ALEPH:1992zwu,ALEPH:1994bih}, we predicted the upper limits on the coupling constants in the $B$-Mesogenesis.

Up to now, the studies on the $B$-Mesogenesis mainly focus on the $B$ decays. However, apart from the generous production of $B$ meson in the LHCb and Belle experiments, $\Lambda_b^0$ is the most baryons that can be produced experimentally. In the $B$-Mesogenesis model, $\Lambda_b^0$ can also undergo baryon number violated decays such as $\Lambda_b\to M \psi$, where $M$ is a neutral pseudoscalar meson including $ \pi^0, K^0, {\bar D}^0 $. Compared with $B\to p/n \psi$, the phase space of $\Lambda_b^0\to M \psi$ is larger,  which can amplify the decay width to a certain extent so that offer more possibility for observing such decays  in the  experiments. In this work,  we will perform a theoretical study on the $\Lambda_b^0\to M \psi$ decays. The decay amplitude will be calculated by LCSR with the use of the light-cone distribution amplitudes (LCDAs) of  $\Lambda_b^0$. One disadvantage of $\Lambda_b^0\to M \psi$ is that since the neutral final pseudoscalar meson makes it  difficult to be detected. In contrast, the charged decay $\Xi_b^{\pm}\to M \psi$ should be a more ideal channel for experimental searching. In the flavor SU(3) symmetry limit, the decay amplitudes of these decay channels are related with each other. The SU(3) symmetry analysis is a powerful tools frequently used in heavy hadron decays \cite{Chau:1986jb,Chau:1987tk,Chau:1990ay,Gronau:1994rj,Zhou:2016jkv,Muller:2015lua,Shi:2017dto}, where it is generally  be  performed in two different frameworks: the topological diagram amplitude (TDA) and the irreducible representation amplitude (IRA) methods. These two methods have been proven to be equivalent in Refs.\cite{Zeppenfeld:1980ex,Gronau:1994rj,Muller:2015lua,He:2018php,He:2018joe,Wang:2020gmn}. In this work, we will choose IRA method to obtain the relations between the decay amplitudes of $\Xi_b^{\pm}\to M \psi$ and those of $\Lambda_b^0\to M \psi$, and use them to predict the decay branching fractions of $\Xi_b^{\pm}\to M \psi$.

This article is organized as follows: Section II introduces the $B$-Mesogenesis scenario proposed by Refs.~\cite{Elor:2018twp,Alonso-Alvarez:2021qfd,Elahi:2021jia}. Section III present a detailed LCSR calculation for the $\Lambda_b^0\to M \psi$ decays. Section IV present a SU(3) analysis on the decay amplitudes, which relates $\Xi_b^{\pm}\to M \psi$ to $\Lambda_b^0\to M \psi$. Section V present numerical calculations on the decay amplitudes and branching fractions. Section VI is a summary of this work.

\section{The $B$-Mesogenesis scenario}
This section gives a brief introduction to the $B$-Mesogenesis scenario \cite{Elor:2018twp,Alonso-Alvarez:2021qfd,Elahi:2021jia}, which aims to simultaneously explain the baryon asymmetry and the existence of dark matter in our Universe. In the $B$-Mesogenesis scenario, the $b$ quark is possible  to decay into two light quarks and a dark baryon $\psi$.  The total baryon number in such decay process is conserved, however, due to the invisible $\psi$ the visible decay products exhibit baryon number non-conserving phenomenon. As proposed by Refs.~\cite{Elor:2018twp,Alonso-Alvarez:2021qfd}, this kind of  baryon number violating decays can emerge from the following two types of effective Lagrangians:
\begin{align}
\mathcal{L}_{\rm eff}^{I}=&-y_{ub}\epsilon_{ijk}Y^{*i}{\bar u}_R^j b_R^{c,k}-y_{cb}\epsilon_{ijk}Y^{*i}{\bar c}_R^j b_R^{c,k}-y_{\psi d}Y_i {\bar \psi}d_R^{c,i}-y_{\psi s}Y_i {\bar \psi}s_R^{c,i}+\rm h.c,\nonumber\\
\mathcal{L}_{\rm eff}^{II}=&-y_{ud}\epsilon_{ijk}Y^{*i}{\bar u}_R^j d_R^{c,k}-y_{us}\epsilon_{ijk}Y^{*i}{\bar u}_R^j s_R^{c,k}-y_{cd}\epsilon_{ijk}Y^{*i}{\bar c}_R^j d_R^{c,k}-y_{cs}\epsilon_{ijk}Y^{*i}{\bar c}_R^j s_R^{c,k}\nonumber\\
&-y_{\psi b}Y_i {\bar \psi}b_R^{c,i}+\rm h.c,\label{eq:effLagran1and2}
\end{align}
which correspond to the type-I and II models in the $B$-Mesogenesis, respectively.  All the quark fields are taken as right handed, the superscript $c$ indicates charge conjugate and the $y$ s are unknown coupling constants. $Y$ is a charged color triplet scalar with $Q_Y=-1/3$, which is assumed to have large mass $M_Y$. In the Type-I model the $b$ quark couples with $u,c$ quarks, the dark anti-baryon $\psi$ couples with $d,s$ quarks. In the Type-II model, the situations of the $b$ and $d,s$ quarks are interchanged. In fact, as proposed by Ref.~\cite{Alonso-Alvarez:2021qfd}, there should be a third type of effective Lagrangian with $Q_Y=2/3$ in $B$-Mesogenesis, which reads as
\begin{align}
\mathcal{L}_{\rm eff}^{III}=&-y_{bd}\epsilon_{ijk}Y^{*i}{\bar b}_R^j d_R^{c,k}-y_{bs}\epsilon_{ijk}Y^{*i}{\bar b}_R^j s_R^{c,k}-y_{\psi u}Y_i {\bar \psi}u_R^{c,i}-y_{\psi c}Y_i {\bar \psi}c_R^{c,i}+\rm h.c.
\end{align}
In this work, we will only consider the case of $Q_Y=-1/3$ to be consistent with the studies of exclusive $B$ meson decay  in $B$-Mesogenesis \cite{Khodjamirian:2022vta,Elor:2022jxy}.  

Integrating out the heavy boson $Y$ in Eq.~(\ref{eq:effLagran1and2}), one can obtain the effective Hamiltonian for the two types of models as:
\begin{align}
&\mathcal{H}_{\rm eff}^{I,uq}=-\frac{y_{ub}y_{\psi q}}{M_Y^2}i\epsilon_{ijk}({\bar \psi} q_R^{c,i})({\bar u}_R^j b_R^{c,k})+{\rm h.c.}=-G_{(uq)}^{I}{\bar {\cal O}}_{(uq)}^{I}\psi^c+{\rm h.c.},\nonumber\\
&\mathcal{H}_{\rm eff}^{II,uq}=-\frac{y_{\psi b}y_{u q}}{M_Y^2}i\epsilon_{ijk}({\bar \psi} b_R^{c,i})({\bar u}_R^j q_R^{c,k})+{\rm h.c.}=-G_{(uq)}^{II}{\bar {\cal O}}_{(uq)}^{II}\psi^c+{\rm h.c.},\label{eq:effHbuq}
\end{align}
where for simplicity, $q=s,d$ and we use $u$ to denote the $u$ or $c$ quarks simultanously. The effective coupling constant and the effective operators in the two models are defined as 
\begin{align}
G_{(uq)}^{I}=\frac{y_{ub}y_{\psi q}}{M_Y^2},~~~{\cal O}_{(uq)}^{I}=-i\epsilon_{ijk}(u^{i T}C P_R b^j)P_R q^k,\nonumber\\
G_{(uq)}^{II}=\frac{y_{\psi b}y_{u q}}{M_Y^2},~~~{\cal O}_{(uq)}^{II}=-i\epsilon_{ijk}(u^{i T}C P_R q^j)P_R b^k,\label{eq:OuqExpression}
\end{align}
where $P_R=(1+\gamma_5)/2$ and $C$ is the charge conjugate matrix.
The baryon number violated decays $\Lambda_b\to \pi \psi$, $\Lambda_b\to K \psi$ and $\Lambda_b\to D \psi$ are induced by $\mathcal{H}_{\rm eff}^{I,ud}$, $\mathcal{H}_{\rm eff}^{I,us}$ and $\mathcal{H}_{\rm eff}^{I,cd}$ respectively.

\section{$\Lambda_b \to M \psi$ decays in LCSR}
\subsection{Hadron level calculation}

\begin{figure}
\centering
\includegraphics[width=0.45\textwidth]{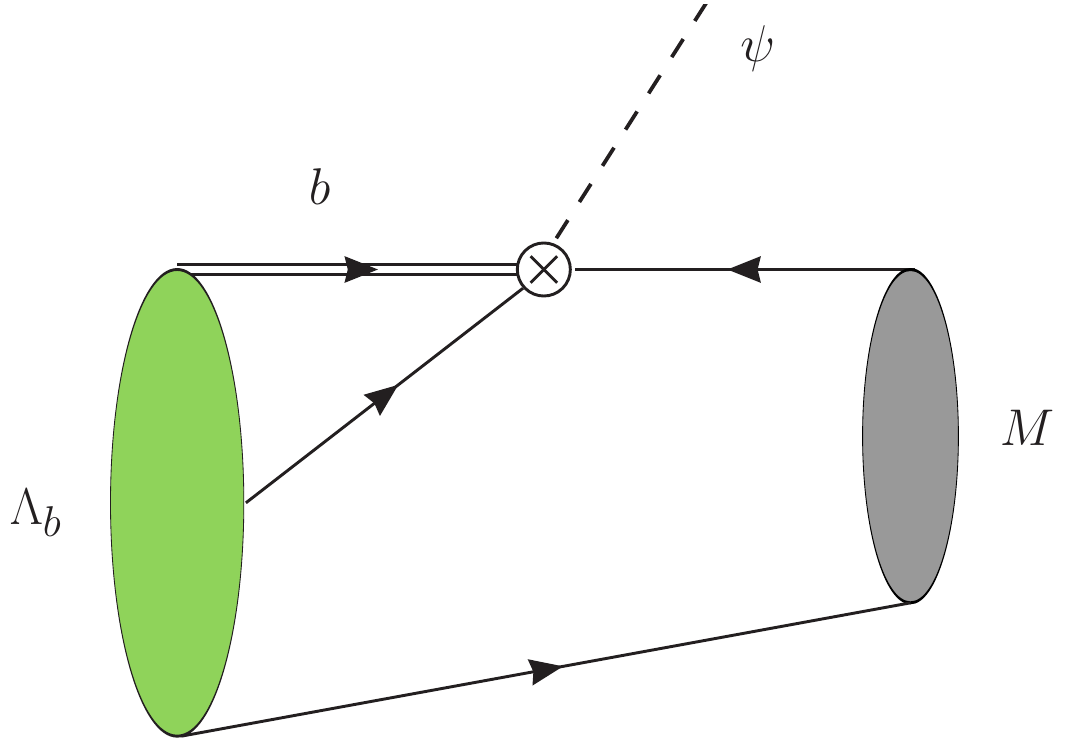}
\caption{The Feynman diagram of the $\Lambda_b \to M \psi$ decays. The green and gray bubbles denote the $\Lambda_b$ and pseudoscalar $M$, respectively,  and the white crossed dot denotes the ${\cal O}_{(uq)}$ vertex.}
\label{fig:LbtoMpsi}
\end{figure} 
In this section, we perform a LCSR calculation on the $\Lambda_b \to M \psi$ decays as shown in Fig.\ref{fig:LbtoMpsi}. The green and gray bubbles denote the $\Lambda_b$ and pseudoscalar $M$, respectively, and the white crossed dot denotes the ${\cal O}_{(uq)}$ vertex. Using the effective Hamiltonians given in Eq.(\ref{eq:effHbuq}), one can  express the decay amplitude as
\begin{align}
i{\cal M}=-G_{(uq)}{\bar u}_{\psi}^c(q, s_{\psi})\langle M(p^{\prime})|{\cal O}_{(uq)}(0)|\Lambda_b(p, s_{\Lambda})\rangle,
\end{align}
where $u_{\psi}$ is the spinor of dark baryon, with momentum $q=p-p^{\prime}$. The transition matrix element on the right hand side above can be parameterized by two form factors:
\begin{align}
\langle M(p^{\prime})|{\cal O}_{(uq)}(0)|\Lambda_b(p, s_{\Lambda})\rangle=P_R \left[F_1(q^2)+\frac{\slashed q}{m_{\Lambda_b}}F_2(q^2)\right] u_{\Lambda_b}(p, s_{\Lambda}).\label{eq:matrixpars}
\end{align}

In the framework of LCSR, the calculation of the transition matrix element given above starts from the following two-point correlation function
\begin{align}
\Pi (p,q)=i\int d^4 x\  e^{ip\cdot x} q^{\mu}   \langle 0|T\{j_{\mu}^M(x) {\cal O}_{(uq)}(0)\}|\Lambda_b(p+q)\rangle,\label{eq:correfunc}
\end{align}
where $j_{\mu}^M$ denotes the interpolation current for the final meson, which  reads as
\begin{align}
j_{\mu}^{\pi}=\frac{1}{\sqrt 2}(\bar u \gamma_{\mu}\gamma_5 u-\bar d \gamma_{\mu}\gamma_5 d),~~~~
j_{\mu}^{K}=\bar s \gamma_{\mu}\gamma_5 d,~~{\text{and}}~~j_{\mu}^{D}=\bar c \gamma_{\mu}\gamma_5 d \label{eq:currents}
\end{align}
for the final states: $M=\pi^0,\ K^0$ and $D^0$ respectively. In LCSR, the correlation function defined in Eq.(\ref{eq:correfunc}) will be calculated both at the hadron and quark-gluon level. The one at hadron level can be expressed by the form factors defined in  Eq.(\ref{eq:matrixpars}). On the other hand, the one at quark-gluon level will be calculated explicitly in QCD, with the light-cone distribution amplitudes (LCDAs) of $\Lambda_b$ taken as non-perturbative inputs. Matching of the  correlation function at these two levels enables us to extract the transition form factors.

At the hadron level, one inserts a complete set of states with the same quantum number of $M$  between the $j_{\mu}^M(q^2)$ current and ${\cal O}_{(uq)}(0)$. The correlation function becomes
\begin{align}
\Pi^H (p,q)=\frac{i f_M}{m_M^2-p^2}(p\cdot q) P_R \left[F_1(q^2)+\frac{\slashed q}{m_{\Lambda_b}}F_2(q^2)\right] u_{\Lambda_b}(p+q)+\int_{s_{\rm th}}^{\infty} ds \frac{\rho^H(s, q)}{s-p^2},\label{eq:hadronCorr1}
\end{align}
where the meson decay constant is defined as: $\langle 0 |j_{\mu}^M(0)|M(p) \rangle=i f_M p_{\mu}$.  Here only the pole contribution from $M$ is expressed explicitly, while the contribution from excited state and continuous spectrum above threshold $s_{\rm th}$ are capsuled in the integration of $\rho^H(s,q)$, which will be suppressed by Borel transformation. Note that in the expression: $p\cdot q=(1/2)(m_{\Lambda_b}^2-q^2-m_M^2)+(1/2)(m_M^2-p^2)$, the second term will cancel the denominator of the pole contribution in Eq.(\ref{eq:hadronCorr1}), so that the terms proportional to it will vanish under the Borel transformation on $p^2$. As a result, the Borel transformed correlation function becomes
\begin{align}
{\cal B}\{\Pi^H\} (p,q)=&\frac{i}{2} f_M (m_{\Lambda_b}^2-q^2-m_M^2) e^{-m_M^2/T^2} P_R \left[F_1(q^2)+\frac{\slashed q}{m_{\Lambda_b}}F_2(q^2)\right] u_{\Lambda_b}(p+q)\nonumber\\
&+\int_{s_{\rm th}}^{\infty} ds\  e^{-s/T^2} \rho^H(s, q),\label{eq:borelhadronCorr1}
\end{align}
where $T$ is the Borel parameter. Now the contributions from excited state and continuous spectrum are suppressed by the exponential term $e^{-s/T^2}$.
Next, the same correlation function will be calculated at the quark-gluon level, where the result can be generally written as a dispersion  integration form
\begin{align}
\Pi^{QCD} (p,q)=\frac{1}{\pi}\int_{s_m}^{\infty} ds \frac{{\rm Im}\Pi^{QCD} (s,q)}{s-p^2},
\end{align}
with $s_m$ being the quark level threshold. From the assumption of quark-hadron duality, the last term of Eq.(\ref{eq:hadronCorr1}), denoting  the excited state and continuous spectrum at hadron level, is equivalent with the spectrum integration above $s_{\rm th}$ at quark-gluon level. After the Borel transformation and subtracting the continuous spectrum contribution, we arrive at the sum rules equation:
\begin{align}
&{\cal B}\{\Pi^H\} (p,q)_{\rm pole}={\cal B}\{\Pi^{QCD}\} (p,q)=\frac{1}{\pi}\int_{s_m}^{s_{\rm th}} ds\  e^{-s/T^2} {\rm Im}\Pi^{QCD} (s,q).\label{eq:sumruleEq}
\end{align}
Thus the form factors $F_{1,2}$ can be extracted through this equation as long as the imaginary part of $\Pi^{QCD}$ is obtained.

\subsection{Quark-gluon level calculation}
\begin{figure}
\centering
\includegraphics[width=0.5\textwidth]{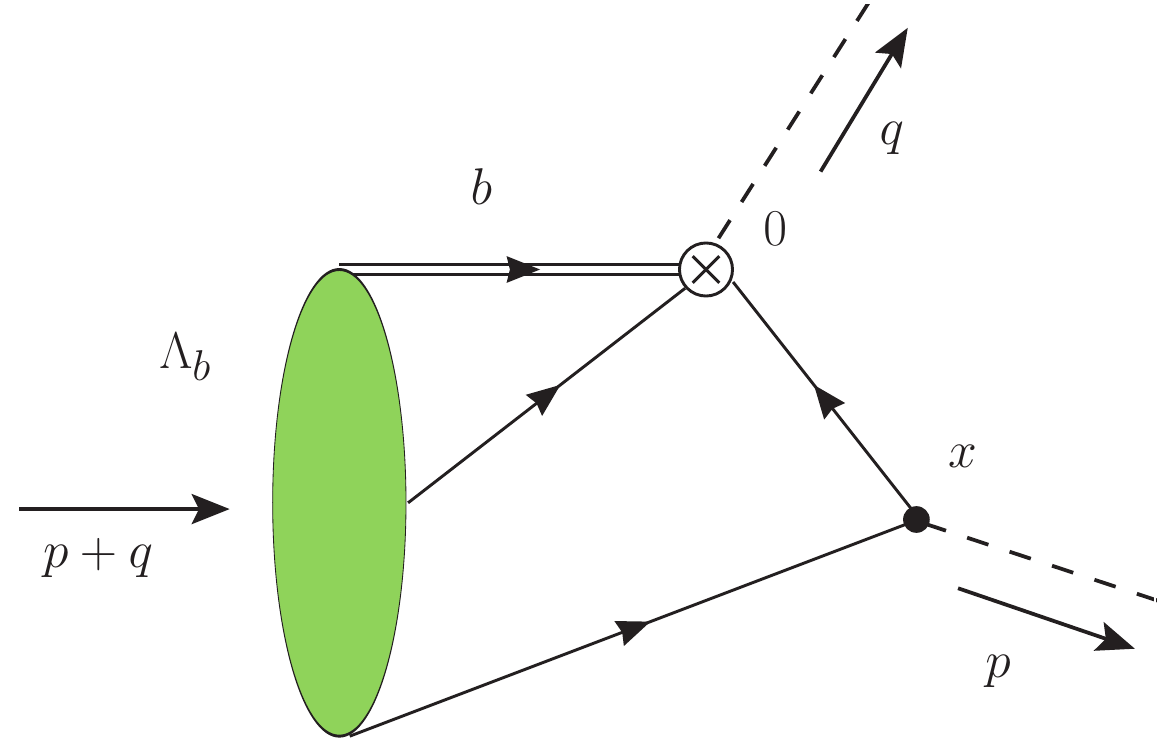}
\caption{The Feynman diagram of the the correlation function defined in Eq.(\ref{eq:correfunc}). The green bubble denotes the $\Lambda_b$ LCDAs,  the white crossed and the black dots denote the ${\cal O}_{(uq)}(x)$ and the $j_{\mu}^M(0)$ vertexes, respectively.}
\label{fig:LbtoMpsiLCSR}
\end{figure}

Now we perform an quark-gluon level calculation for the correlation function defined in Eq.(\ref{eq:correfunc}). Here we take the type-I model and the case of $M={\bar D}^0$ as an example to illustrate the calculation. Using the quark level expression of $j_{\mu}^D$ given in Eq.(\ref{eq:currents}), one can write the correlation function as
\begin{align}
\Pi_{\Lambda_b\to D}^{QCD} (p,q)=&-\epsilon_{ijk}\int d^4 x e^{i p\cdot x} \left[P_R C S_{c}(0,x)\slashed q \gamma_5\right]_{\gamma\alpha}(P_R)_{\tau\beta}\nonumber\\
&\times\langle 0|u_{\alpha}^i(q^2) d_{\beta}^j(0)b_{\gamma}^k(0)|\Lambda_b(p+q)\rangle,
\end{align}
where $S_c$ is the charm quark propagator. 
The non-local three quark matrix element above is expressed by the $\Lambda_b$ LCDAs, which are defined as \cite{Ball:2008fw,Duan:2022uzm}
\begin{align}
&\langle 0|u_{\alpha}^i(x_1) d_{\beta}^j(x_2)b_{\gamma}^k(0)|\Lambda_b(v)\rangle\nonumber\\
=&\frac{1}{8}f_{\Lambda_b}^{(2)}\Psi_2(t_1,t_2)(\slashed{\overline{n}}\gamma_5C)_{\alpha\beta}u_{\Lambda_b\gamma}(v)+\frac{1}{4}f_{\Lambda_b}^{(1)}\Psi_3^s(t_1,t_2)(\gamma_5C)_{\alpha\beta}u_{\Lambda_b\gamma}(v) \notag \\
   &-\frac{1}{8}f_{\Lambda_b}^{(1)}\Psi_3^\sigma(t_1,t_2)(i\sigma_{\overline{n}n}\gamma_5C)_{\alpha\beta}u_{\Lambda_b\gamma}(v)+\frac{1}{8}f_{\Lambda_b}^{(2)}\Psi_4(t_1,t_2)(\slashed{n}\gamma_5C)_{\alpha\beta}u_{\Lambda_b\gamma}(v), \label{eq:LCDAs}
\end{align}
where $\sigma_{\overline{n}n}=\sigma_{\mu\nu}\overline{n}^{\mu}n^{\nu}$ and $x_1=t_1 n, x_2=t_2 n$ are on the light cone. Inversely, the light cone vectors $n, \bar n$ can be expressed by the coordinates as  
   \begin{gather*}
   	n_\mu=\frac{x_\mu}{v\cdot x},\quad \overline{n}_\mu=2v_\mu-\frac{x_\mu}{v\cdot x}.
   \end{gather*}
At the quark-gluon level, the corresponding diagram for the correlation function defined in Eq.(\ref{eq:correfunc}) is shown in Fig.\ref{fig:LbtoMpsiLCSR}. The green bubble denotes the $\Lambda_b$ LCDAs, the white crossed and the black dots denote the ${\cal O}_{(uq)}(x)$ at and the $j_{\mu}^M(0)$ vertexes, respectively.
   
 The $\Psi_2, \Psi_3^s, \Psi_3^\sigma, \Psi_4$ in Eq.(\ref{eq:LCDAs}) are the LCDAs with different twists. In the momentum space, the LCDAs can be characterized  by the total momentum $\omega$ of the two light quarks in $\Lambda_b$, and the momentum fraction $u$ of one light quark:
   \begin{equation}
   \Psi_i(t_1,t_2)=\int_{0}^{\infty}\omega d\omega\int_{0}^{1}due^{-i\omega(t_1u+t_2\overline{u})}{\psi}_i(\omega,u),
   \end{equation}
   with $\overline{u}=1-u$, and  \cite{Ball:2008fw}
   \begin{align}
   {\psi}_2(\omega,u)=&\omega^2u(1-u)[\frac{1}{\epsilon_0^4}e^{-\omega/\epsilon_0}+a_2C_2^{3/2}(2u-1)\frac{1}{\epsilon_1^4}e^{-\omega/\epsilon_1}],  \nonumber\\
   {\psi}_3^s(\omega,u)=&\frac{\omega}{2\epsilon_3^3}e^{-\omega/\epsilon_3}, \nonumber\\
   {\psi}_3^\sigma(\omega,u)=&\frac{\omega}{2\epsilon_3^3}(2u-1)e^{-\omega/\epsilon_3},\nonumber\\
   {\psi}_4(\omega,u)=&5\mathcal{N}^{-1}\int_{\omega/2}^{s_0^{\Lambda_b}}dse^{-s/\tau}(s-\omega/2)^3, \label{eq:fourLCDAs}
   \end{align}
where $C_2^{3/2}(2u-1)$ is the Gegenbauer polynomial, and the normalization factor $\mathcal{N}$ in $\tilde{\psi}_4(\omega,u)$ reads as $\mathcal{N}=\int_0^{s_0^{\Lambda_b}} ds s^5 e^{-s/\tau}$.  The parameters in these four LCDAs are $\epsilon_0=200^{+130}_{-60}$ $\rm{MeV}, \epsilon_1=650^{+650}_{-300}$ $\rm{MeV}, \epsilon_3=230$ $\rm{MeV}$, $a_2=0.333^{+0.250}_{-0.333}$, $s_0^{\Lambda_b}=1.2$ GeV and $\tau=0.4\sim 0.8$ GeV. \cite{Ball:2008fw}. 

Using the LCDAs defined in Eq.(\ref{eq:LCDAs}), and the free charm quark propagator,  one can express the correlation function as a convolution of the perturbative part and the LCDAs. Firstly, the twist-3s contribution can be straightforwardly written as  
 \begin{align}
\Pi_{\Lambda_b\to D}^{QCD(3s)} (p,q)=&\frac{i}{4}f_{\Lambda_b}^{(1)}\int _0^{2 s_0^{\Lambda_b}}d\omega\ \omega\int_0^1 du\ \psi_{3s}(\omega,u)\frac{1}{(u\omega v-p)^2-m_c^2} P_{R}\slashed q (u\omega \slashed v - \slashed p) u_{\Lambda_b}(v).
\end{align}
To extract the imaginary part of correlation function, one has to calculate its discontinuity across the complex plane of $p^2$. The denominator can be rewritten as
\begin{align}
&(u\omega v-p)^2-m_c^2=\left(1-\frac{u\omega}{m_{\Lambda_b}}\right)(p^2-\Sigma_c),\nonumber\\
&\Sigma_c=\frac{1}{m_{\Lambda_b}-u\omega}\left[u\omega (m_{\Lambda_b}^2-q^2)+m_{\Lambda_b}(m_c^2-u^2 \omega^2)\right].
\end{align}
Now the imaginary part comes from the term
\begin{align}
{\rm Im}\left[\frac{1}{(u\omega v-p)^2-m_c^2}\right]=\frac{1}{2i}(-2\pi i)\frac{m_{\Lambda_b}}{m_{\Lambda_b}-u\omega}\delta(s-\Sigma_c),\label{eq:deltaFunc}
\end{align}
where $s=p^2$. Expressing $\Pi_{\Lambda_b\to D}^{QCD(3s)}$ as a dispersion integration in $s_{m}<s<s_{\rm th}$, performing the Borel transformation, and integrating out $s$ by the delta function in Eq.(\ref{eq:deltaFunc}), one arrives at
\begin{align}
{\cal B}\{\Pi_{\Lambda_b\to D}^{QCD(3s)}\}(T,q)=&\frac{i}{4}f_{\Lambda_b}^{(1)}\int _0^{2 s_0^{\Lambda_b}}d\omega \int_0^1 du\ \theta(\Sigma_c-s_m)\theta(s_{\rm th}-\Sigma_c)e^{-\Sigma_c/T^2}\nonumber\\
&\times \psi_{3s}(\omega,u) P_R \left[m_{\Lambda_b}\omega \slashed q-\frac{m_{\Lambda_b}\omega}{m_{\Lambda_b}-u\omega}q^2\right].
\end{align}

The calculation for the twist-3$\sigma$ contribution is more involved. The corresponding correlation function can be firstly written as
 \begin{align}
\Pi_{\Lambda_b\to D}^{QCD(3\sigma)} (p,q)=&-\frac{i}{4}f_{\Lambda_b}^{(1)}\int d^4 x\ e^{ip\cdot x}\int _0^{2 s_0^{\Lambda_b}}d\omega\ \omega\int_0^1 du\ e^{-i u \omega v\cdot x} \int\frac{d^4 k}{(2\pi)^4} e^{i k\cdot x}\nonumber\\
&\times \psi_{3s}(\omega,u) P_{R}(1-\slashed n \slashed v)\slashed q \frac{1}{\slashed k}u_{\Lambda_b}(v),\label{eq:3sigmaCorr}
\end{align}
where the 4-velocity of $\Lambda_b$ is $m_{\Lambda_b}v=p+q$. It can be found that the term containing no $\slashed n$ above has almost the same form as that of $\Pi_{\Lambda_b\to D}^{QCD(3s)}$. The corresponding contribution to the Borel transformed $\Pi_{\Lambda_b\to D}^{QCD(3\sigma)}$ is
\begin{align}
{\cal B}\{\Pi_{\Lambda_b\to D}^{QCD(3\sigma(1))}\}(T,q)=&-\frac{i}{4}f_{\Lambda_b}^{(1)}\int _0^{2 s_0^{\Lambda_b}}d\omega \int_0^1 du\ e^{-\Sigma_c/T^2}\theta(\Sigma_c-s_m)\theta(s_{\rm th}-\Sigma_c)\nonumber\\
&\times \psi_{3\sigma}(\omega,u) P_R \left[m_{\Lambda_b}\omega \slashed q-\frac{m_{\Lambda_b}\omega}{m_{\Lambda_b}-u\omega}q^2\right].
\end{align}

Next we consider the term proportional to $\slashed n$ in Eq.(\ref{eq:3sigmaCorr}). Note that $n_{\mu}=x_{\mu}/v\cdot x$, to eliminate the $1/v\cdot x$ one can define modified LCDAs as
\begin{align}
{\bar \psi}_{i}(\omega, u)=\int_0^{\omega}d\tau \tau \psi_i(\tau, u),
\end{align}
with $i=2,3s,3\sigma,4$. Thus the $1/v\cdot x$ can be eliminated through integration by part:
\begin{align}
\int_0^{2 s_0^{\Lambda_b}}  d\omega\  \omega \psi_i(\omega,u)e^{-iu\omega v\cdot x}\frac{x_{\mu}}{v\cdot x}=i u \int_0^{2 s_0^{\Lambda_b}} d\omega {\bar \psi}_{i}(\omega, u)e^{-iu\omega v\cdot x} x_{\mu} .
\end{align}
Here the boundary term has been omitted since large $\omega$ in the exponential induces high frequency oscillation under the integration of $x$, which suppresses its contribution. The $x_{\mu}$ can be written as $-i \partial/\partial p^{\mu}$, and thus the $\slashed n$ term in Eq.(\ref{eq:3sigmaCorr}) contributes:
 \begin{align}
\Pi_{\Lambda_b\to D}^{QCD(3\sigma(n))} (p,q)=&\frac{i}{4}f_{\Lambda_b}^{(1)}\int _0^{2 s_0^{\Lambda_b}}d\omega\ \omega\int_0^1 du \ {\bar \psi}_{3\sigma}(\omega,u)\nonumber\\
&\times P_R \frac{\partial}{\partial p^{\mu}} \left[\frac{1}{(p-u\omega)^2-m_c^2}\gamma^{\mu}\slashed v \slashed q (u\omega \slashed v-\slashed p)\right]u_{\Lambda_b}(v).
\end{align}
After dispersion integration, performing the Borel transformation, and introducing auxiliary mass to lower the higher power of denominators: $1/(p^2-\Sigma_c)^2=(\partial/\partial M^2)[1/(p^2-\Sigma_c-M^2)]|_{M^2=0}$, one obtains the $\slashed n$ term contribution as
\begin{align}
&{\cal B}\{\Pi_{\Lambda_b\to D}^{QCD(3\sigma(n))}\}(T,q)\nonumber\\
=&-\frac{i}{2}f_{\Lambda_b}^{(1)}\frac{\partial}{\partial M^2}\int _0^{2 s_0^{\Lambda_b}}d\omega \int_0^1 du\ u\  {\bar \psi}_{3\sigma}(\omega,u) \theta[(\Sigma_c+M^2)-s_m]\theta[s_{\rm th}-(M^2+\Sigma_c)] e^{-(\Sigma_c+M^2)/T^2}\nonumber\\
&\times  P_R \left(\frac{m_{\Lambda_b}}{m_{\Lambda_b}-u\omega}\right)^2 \left[A(p^2,q^2)+(2v\cdot q)m_c^2+B(p^2,q^2)\slashed q \right]u_{\Lambda_b}(v){\Big |}_{M^2=0,\  p^2=\Sigma_c+M^2}\ ,
\end{align}
where 
\begin{align}
A(p^2,q^2)&=(2 v\cdot q)(m_{\Lambda_b}-u\omega)(m_{\Lambda_b}+u\omega-2v\cdot p)-2(m_{\Lambda_b}-v\cdot p)q^2,\nonumber\\
B(p^2,q^2)&=q^2-(m_{\Lambda_b}-u\omega)(m_{\Lambda_b}+u\omega-2v\cdot p),
\end{align}
and note that all the Lorentz invariants involved above should be expressed by $p^2, q^2$ as
\begin{align}
v\cdot q&=\frac{1}{2m_{\Lambda_b}}(m_{\Lambda_b}^2+q^2-p^2),~~~
v\cdot p=\frac{1}{2m_{\Lambda_b}}(m_{\Lambda_b}^2+p^2-q^2),\nonumber\\
p\cdot q&=\frac{1}{2}(m_{\Lambda_b}^2-q^2-p^2)
\end{align}

The twist-2 and 4 contributions to the correlation function can be obtained similarly as that of twist-3. The corresponding Borel transformed correlation function is
\begin{align}
&{\cal B}\{\Pi_{\Lambda_b\to D}^{QCD(2)+(4)}\}(p,q)\nonumber\\
=&\frac{i}{4}m_c f_{\Lambda_b}^{(2)}\int _0^{2 s_0^{\Lambda_b}}d\omega \int_0^1 du\ u\ e^{-\Sigma_c/T^2}\theta(\Sigma_c-s_m)\theta(s_{\rm th}-\Sigma_c)\nonumber\\
&\times {\psi}_{2}(\omega,u) \left(\frac{\omega m_{\Lambda_b}}{m_{\Lambda_b}-u\omega}\right) P_R  \left[2v\cdot q- \slashed q \right]u_{\Lambda_b}(v)|_{M^2=0}\nonumber\\
+&\frac{i}{4}m_c f_{\Lambda_b}^{(2)}\frac{\partial}{\partial M^2}\int _0^{2 s_0^{\Lambda_b}}d\omega \int_0^1 du\ u\ e^{-(\Sigma_c+M^2)/T^2}\theta(\Sigma_c+M^2-s_m)\theta(s_{\rm th}-M^2-\Sigma_c)({\bar \psi}_{2}-{\bar \psi}_{4})(\omega,u)\nonumber\\
&\times \left(\frac{m_{\Lambda_b}}{m_{\Lambda_b}-u\omega}\right)^2 P_R \left[2 p\cdot q + q^2-2 u \omega v\cdot q -(m_{\Lambda_b}-u \omega) \slashed q \right]u_{\Lambda_b}(v){\Big |}_{M^2=0,\  p^2=\Sigma_c+M^2}\ .
\end{align}
It can be found that here the twist-2 and 4 contributions are proportional to $m_c$. In the case of $\Lambda_b \to \pi, K + \psi$ decays, the twist-2 and 4 contributions are proportional to $m_u=m_d=0$ and thus vanish. For the $\Lambda_b \to \pi, K + \psi$ decays in the type-I model, the calculation of the correlation function in Eq.(\ref{eq:correfunc}) is almost the same, which can be found in the Appendix \ref{app:analyI}.

Generally, as Eq.(\ref{eq:borelhadronCorr1}) shows, the correlation function has two independent spinor structures: $1$ and $\slashed q$. In the type-I model both of them exists in the  correlation function. However, it can be found that the $\slashed q$ term is absent in the type-II model. As a result, only $F_1$ contributes to the $\Lambda_b^0 \to \pi^0, K^0, {\bar D}^0  + \psi$ decays, while $F_2$ vanishes up to the twist-3 LCDA contributions.  On the other hand, in the type-II model the $\Lambda_b^0 \to \pi^0 + \psi$ decay is forbidden due to the flavor SU(3) limit, which will be discussed in the next section.
The corresponding analytical results in the type-II models are given in the Appendix \ref{app:analyII}.

\section{SU(3) analysis}\label{sec:SU3}
In the above study we have only focused on the decay processes $\Lambda_b^0 \to \pi^0, K^0, {\bar D}  + \psi$, while the various decay channels $\Xi_b^{0,-}\to \pi^{0,-}, K^{0,-} + \psi$ and $\Lambda_b^0, \Xi_b^{0} \to \eta  + \psi$ have not been considered. In principle, the amplitudes of these missed processes can also be calculated as soon as the LCDAs of $\Xi_b$ or $\eta$ are known, which have not been studied as well as the LCDAs of $\Lambda_b$. However, with the use of flavor SU(3) symmetry, one can still able to predict the decay widths of these missed channels as long as the decay widths of $\Lambda_b^0 \to \pi^0, K^0 + \psi$ are known. 

As shown in Eq.(\ref{eq:effHbuq}), the effective Hamiltonian: $-G_{uq} {\cal O}_{(uq)}$ contains two light quark fields $u,\ d$
or $u,\ s$. In the flavor SU(3) representation, this Hamiltonian can be  represented
 by a rank-two tensor $H_{ij}$, where $i ,j$ are the flavor indexes. Its non-vanishing
components are $H_{12}=G_{ud}$ and $H_{13}=G_{us}$
for $b\to u,d$ and $b\to u,s$ transitions, respectively.  Note that $H_{ij}$ is reducible so that can be  further 
be reduced into three SU(3) irreducible representations: a symmetric and traceless tensor $\bar{S}_{\{ij\}}$, an anti-symmetry tensor $T_{[ij]}$ and a trace term $I \delta_{ij}$:
\begin{align}
H_{ij} & =\bar{S}_{\{ij\}}+T_{[ij]}+I_{\{ij\}},
\end{align}
where
\begin{align}
\bar{S}_{\{ij\}} =\frac{1}{2}H_{ij}+\frac{1}{2}H_{ji}-\frac{1}{3}\delta_{ij}H_{kk},~~~
T_{[ij]} =\frac{1}{2}H_{ij}-\frac{1}{2}H_{ji}.
\end{align}
The non-vanishing components are $\bar{S}_{12}=\bar{S}_{21}=\frac{1}{2}$,
$T_{12}=-T_{21}=\frac{1}{2}$ for $b\to u,d$,  and $\bar{S}_{13}=\bar{S}_{31}=\frac{1}{2}$,
$T_{13}=-T_{31}=\frac{1}{2}$ for $b\to u,s$. It should be mentioned that in the type-II model, as shown by Eq.\ref{eq:OuqExpression} the two light flavors are anti-symmetrized, so that $\bar{S}_{ij}$ vanishes in this case.

On the other hand, the initial anti-triplet baryons $T_{\bf{b\bar 3}}$ and the final light mesons $M$ can also be expressed by SU(3) irreducible representations as
\begin{eqnarray}
 (T_{\bf{b\bar 3}}^{ij})= \left(\begin{array}{ccc} 0 & \Lambda_b^0  &  \Xi_b^0  \\ -\Lambda_b^0 & 0 & \Xi_b^- \\ -\Xi_b^0   &  -\Xi_b^-  & 0
  \end{array} \right)_{ij}, \;\;\;  M^{i}_j=\begin{pmatrix}
 \frac{\pi^0}{\sqrt{2}}+\frac{\eta}{\sqrt{6}}  &\pi^+ & K^{+}\\
 \pi^-& -\frac{\pi^0}{\sqrt{2}}+\frac{\eta}{\sqrt{6}} &{K^{0}}\\
 K^{-}&\overline K^{0} & -\frac{2\eta}{\sqrt{6}}
 \end{pmatrix} _{ij}. 
\end{eqnarray}
Using $T_{\bf{b\bar 3}}$, $M^{i}_j$, $\bar{S}_{ij}$ and $T_{ij}$, one can construct SU(3) invariant amplitude as
\begin{align}
{\cal A}=s_1\  T_{\bf{b\bar 3}}^{ij} {\bar S}_{jk} (M^T)^k_i \psi+t_1\  T_{\bf{b\bar 3}}^{ij} T_{jk} (M^T)^k_i \psi+t_2\  T_{\bf{b\bar 3}}^{ij} T_{ij} (M^T)^k_k \psi.\label{eq:SU3amp}
\end{align}
Here we have not considered the color singlet $\eta_1$ in the construction of $M^i_j$, thus the third term above vanishes due to $M^k_k=0$. The unknown amplitudes $s_1, t_1, t_2$ contain all the information about the strong interaction in the decays. All the decay amplitudes of anti-triplet bottom baryon decays into a light meson and dark baryon are listed in Table \ref{Tab:SU3amp}, where the left and right two columns correspond to $b\to d$ and $b\to s$ transitions. 
\begin{table}
  \caption{Decay amplitudes of anti-triplet bottom baryon decays into a light meson and dark baryon from Eq.(\ref{eq:SU3amp}), where the left and right columns correspond to $b\to d$ and $b\to s$ transitions, respectively.}
\label{Tab:SU3amp}
\begin{tabular}{|c|c|c|c|}
\hline
\hline
Channel & Amplitude & Channel & Amplitude \tabularnewline 
\hline 
$\Lambda_b^0 \to \pi^0 \psi$ & $\sqrt{2} s_1 G_{ud}$ & $\Lambda_b^0 \to K^0 \psi$ & $- (s_1+t_1) G_{us}$ \tabularnewline 
$\Xi_b^0 \to {\bar K}^0  \psi$ & $-(s_1+t_1) G_{ud}$ & $\Xi_b^0 \to \pi^0  \psi$ & $\frac{1}{\sqrt{2}}(s_1-t_1) G_{us}$ \tabularnewline 
$\Xi_b^- \to K^-  \psi$ & $-(s_1-t_1) G_{ud}$ & $\Xi_b^- \to \pi^-  \psi$ & $(s_1-t_1) G_{us}$ \tabularnewline 
$\Lambda_b^0 \to \eta  \psi$ & $-\sqrt{\frac{2}{3}} t_1 G_{ud}$ & $\Xi_b^0 \to \eta  \psi$ & $\Big(\frac{1}{\sqrt{6}} t_1-\sqrt{\frac{3}{2}} s_1\Big) G_{us}$ \tabularnewline 
\hline 
\end{tabular}
\end{table}
Note that the amplitude of $\Lambda_b^0 \to \pi^0 + \psi$ is proportional to $s_1$ which is absent in the type-II model. Therefore, in the type-II model the $\Lambda_b^0 \to \pi^0 + \psi$ decay is forbidden, and we have $s_1=0$ in Table \ref{Tab:SU3amp} so that the amplitudes of all the decay channels are proportional to $t_1$. On the other hand, in the type-I model, using the decay amplitudes of $\Lambda_b^0 \to \pi^0 \psi$ and $\Lambda_b^0 \to K^0 \psi$ calculated in this work by LCSR, we can determine  the fraction $\xi_{K/\pi}=t_1/s_1$ as
\begin{align}
-\frac{1}{\sqrt 2}(1+\xi_{K/\pi})\lambda_{s/d}=\frac{{\cal A}(\Lambda_b^0 \to K^0 \psi)}{{\cal A}(\Lambda_b^0 \to \pi^0 \psi)},\label{eq:defofxi}
\end{align}
with $\lambda_{s/d}=G_{us}/G_{ud}$. This enables us to predict all the decay amplitudes listed in Table \ref{Tab:SU3amp}.

\section{Numerical Results}

The hadron masses are taken as: $m_{\Lambda_b}=5.62$ GeV, $m_{\pi}=0.135$ GeV, $m_{K}=0.498$ GeV and  $m_{D}=1.86$ GeV. The quark masses are taken as $m_u=m_d=0$ and $m_c=1.1$ GeV at $\mu=3$ GeV \cite{ParticleDataGroup:2022pth}. $\tau=0.6$ GeV is taken as the center value in its range $\tau=0.4\sim 0.8$ GeV.  The threshold parameters should be above the lowest state while nearly below the next lowest state. The next lowest states corresponding to $\pi, D, K$ are $\pi(1300), K(1460), D(2550)$ \cite{ParticleDataGroup:2022pth} and $s_{\rm th}$ can be parameterized  as
\begin{align}
s_{\rm th}^{\pi, K, D}=(1-\lambda)m_{\pi, K, D}^2 + \lambda\   m_{\pi(1300), K(1460), D(2550)}^2,
\end{align}
where $\tau=0$ and $\tau=1$ correspond to the lowest and the next lowest states, respectively. In this work, to estimate the uncertainties from the threshold parameter,  we choose the range $0.6<\lambda <1.0$ for error analysis in numerical calculations.

\begin{figure}
\centering
\includegraphics[width=1.0\textwidth]{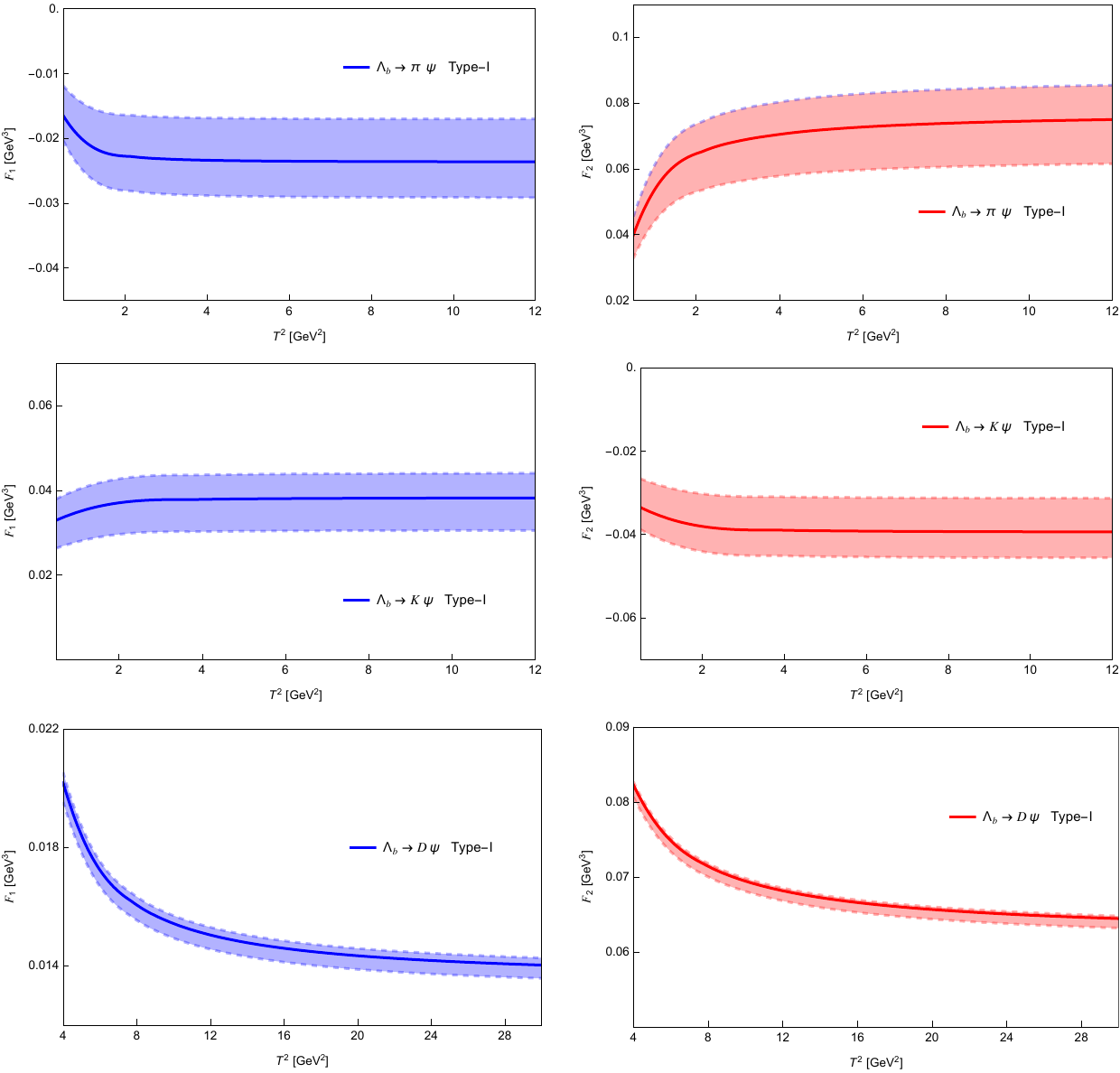}
\caption{The form factors $F_1, F_2$ as functions of $T^2$ in the type-I model, with $q^2=0$. The band width denotes the uncertainty of $s_{\rm th}$: $0.6<\lambda <1.0$.}
\label{fig:FFvsT2typeI}
\end{figure} 
\begin{figure}
\centering
\includegraphics[width=1.0\textwidth]{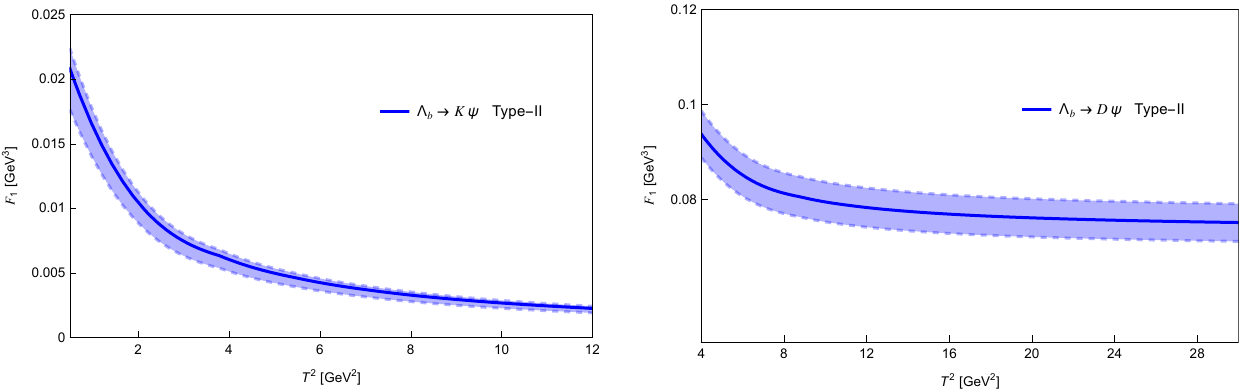}
\caption{The form factors $F_1, F_2$ as functions of $T^2$ in the type-II model, with $q^2=0$. The band width denotes the uncertainty of $s_{\rm th}$: $0.6<\lambda <1.0$.}
\label{fig:FFvsT2typeII}
\end{figure} 
The form factors $F_1, F_2$ as functions of $T^2$ are shown in Fig.\ref{fig:FFvsT2typeI} and Fig.\ref{fig:FFvsT2typeII} for the type-I and II models respectively, where $q^2=0$. In principle, the form factors should be independent of the unphysical  Borel parameter. Therefore, one has to choose a window for $T^2$ where the behavior of $F_1, F_2$ are almost stable. From Fig.\ref{fig:FFvsT2typeI}, it can be found that the stable regions are at large $T^2$, and the $T^2$ windows are chosen as
\begin{align}
\Lambda_b\to \pi \psi,\ K \psi:&~~~4\  {\rm GeV}^2<T^2<12\  {\rm GeV}^2,\nonumber\\
\Lambda_b\to D \psi:&~~~12\  {\rm GeV}^2<T^2<28\  {\rm GeV}^2
\end{align}
for the both type-I and II models. 

\begin{figure}
\centering
\includegraphics[width=1.0\textwidth]{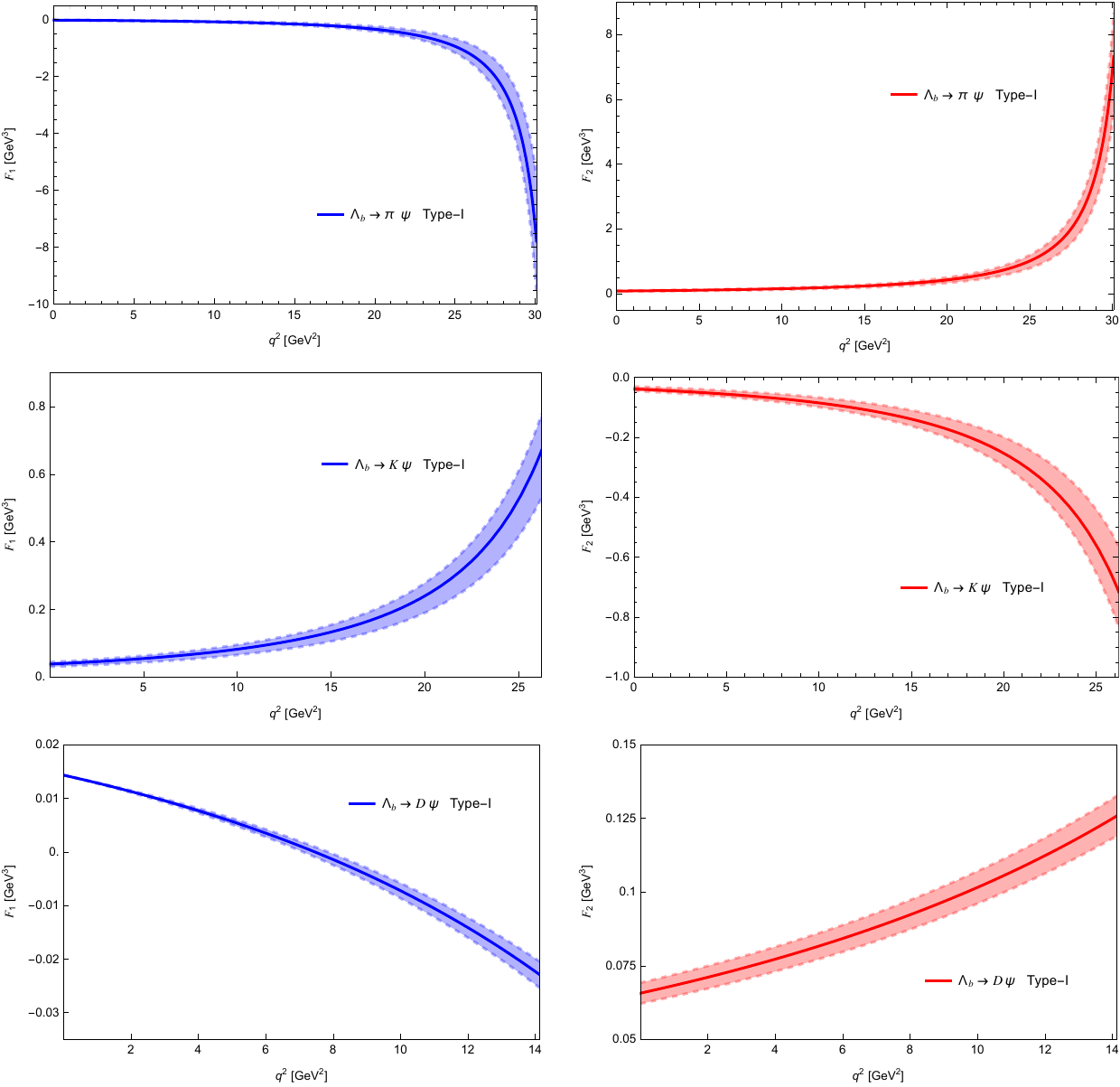}
\caption{The form factors $F_1, F_2$ as functions of $q^2$ in the type-I model, with $q^2=0$. The band width denotes the combined  uncertainties of the Borel parameter and $s_{\rm th}$: $0.6<\lambda <1.0$.}
\label{fig:FFvsq2typeI}
\end{figure}

\begin{figure}
\centering
\includegraphics[width=1.0\textwidth]{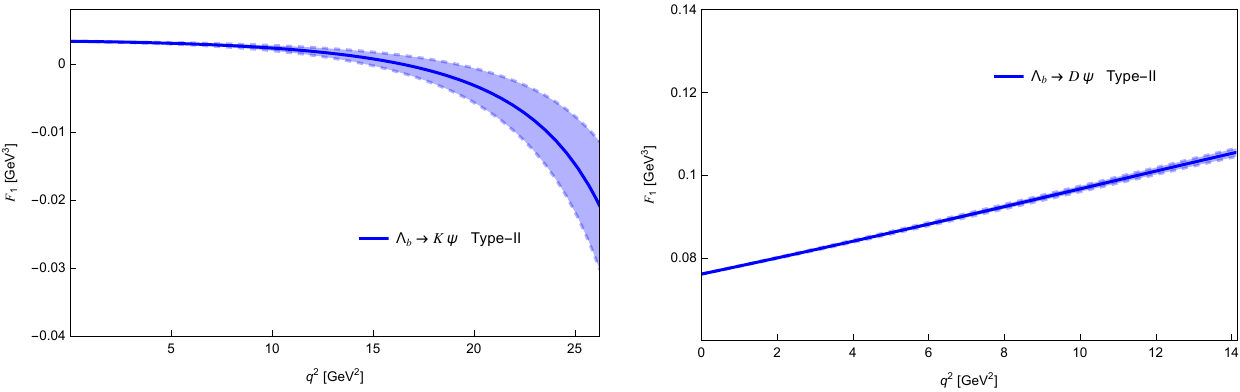}
\caption{The form factors $F_1, F_2$ as functions of $q^2$ in the type-II model, with $q^2=0$. The band width denotes the combined  uncertainties of the Borel parameter and $s_{\rm th}$: $0.6<\lambda <1.0$.}
\label{fig:FFvsq2typeII}
\end{figure}

On the other hand, note that the $F_{1,2}(q^2)$ obtained by LCSR are only reliable in the small positive $q^2$ or $q^2<0$ regions, instead of the physical region $q^2>0$. Therefore, one has to fit the form factors in the $q^2<0$ region by a suitable parameterization function, and extent the form factors to physical region. In this work, we use the z-series formula \cite{Bourrely:2008za} with single pole structure to perform the fitting, which reads as
\begin{align}
F_{1,2}(q^2)=\frac{F_{1,2}(0)}{1-\frac{q^2}{m_{\rm pole}^2}}\left[1+ b_{1,2} (z(q^2)-z(0))+c_{1,2}(z(q^2)-z(0))^2\right],\label{eq:fitformula}
\end{align}
where $m_{\rm pole}$ is the mass of the lowest baryon state that can be created by ${\cal O}_{uq}^{\dagger}$ from the vaccum. For the $\Lambda_b\to \pi, K, D$ transitions, $m_{\rm pole}$ is chosen as $m_{\Lambda_b},  m_{\Xi_b}=5.79 \text{GeV}, m_{\Xi_{bc}}=6.94 \text{GeV}$ respectively, where $m_{\Xi_{bc}}$ is taken from QCD sum rules calculation \cite{Hu:2017dzi}.  The $F_{1,2}(0), b_{1,2}$ and  $c_{1,2}$ in Eq.(\ref{eq:fitformula}) are three fitting parameters, and the $z$ function is defined as
\begin{align}
z(q^2)&=\frac{\sqrt{t^+ -q^2}-\sqrt{t^+ -t_0}}{\sqrt{t^+ -q^2}+\sqrt{t^+ -t_0}}
\end{align}
with $t^{\pm}=(m_{\Lambda_b}\pm m_M)^2$ and $t_0=t^{+}\left(1-\sqrt{1-t^-/t^+}\right)$. The fitting results of $F_{1,2}(0), b_{1,2}$ and  $c_{1,2}$ are listed in Table \ref{Tab:fitpars}, while the form factors extended to the physical region are shown in Fig.\ref{fig:FFvsq2typeI} and Fig.\ref{fig:FFvsq2typeII}  for the type I and II models, respectively.

\begin{table}
  \caption{Fitting parameters of the $F_{1,2}$ for $\Lambda_b \to \pi, K, D$ transitions in the Type-I and II models.}
\label{Tab:fitpars}
\begin{tabular}{|c|ccc|ccc|}
\hline
\hline
Type-I & $F_1(0)$ & $b_1$ & $c_1$ & $F_2(0)$ & $b_2$ & $c_2$ \tabularnewline
\hline 
$\Lambda_b^0 \to \pi^0$ & $-0.02\pm 0.005$ & $-15.77$ & 16.41 & $0.07\pm 0.014$ & -3.81 & 4.43 \tabularnewline
\hline 
$\Lambda_b^0 \to K^0$ & $0.04\pm 0.006$ & -5.74 & 12.34 & $-0.04\pm 0.006$ & -5.88 & 13.32 \tabularnewline
\hline 
$\Lambda_b^0 \to {\bar D}^0$ & $0.01\pm 0.001$ & 26.79 & -35.06 & $0.06\pm 0.003$ & -4.0 & 11.89 \tabularnewline
\hline 
Type-II & $F_1(0)$ & $b_1$ & $c_1$ & $F_2(0)$ & $b_2$ & $c_2$ \tabularnewline
\hline 
$\Lambda_b^0 \to K^0$ & $0.003\pm 0.001$ & 6.21 & -5.76 & -- & -- & -- \tabularnewline
\hline 
$\Lambda_b^0 \to {\bar D}^0$ & $0.07\pm 0.0$ & -1.01 & -17.6 & -- & -- & -- \tabularnewline
\hline 
\end{tabular}
\end{table}

\begin{figure}
\centering
\includegraphics[width=0.48\textwidth]{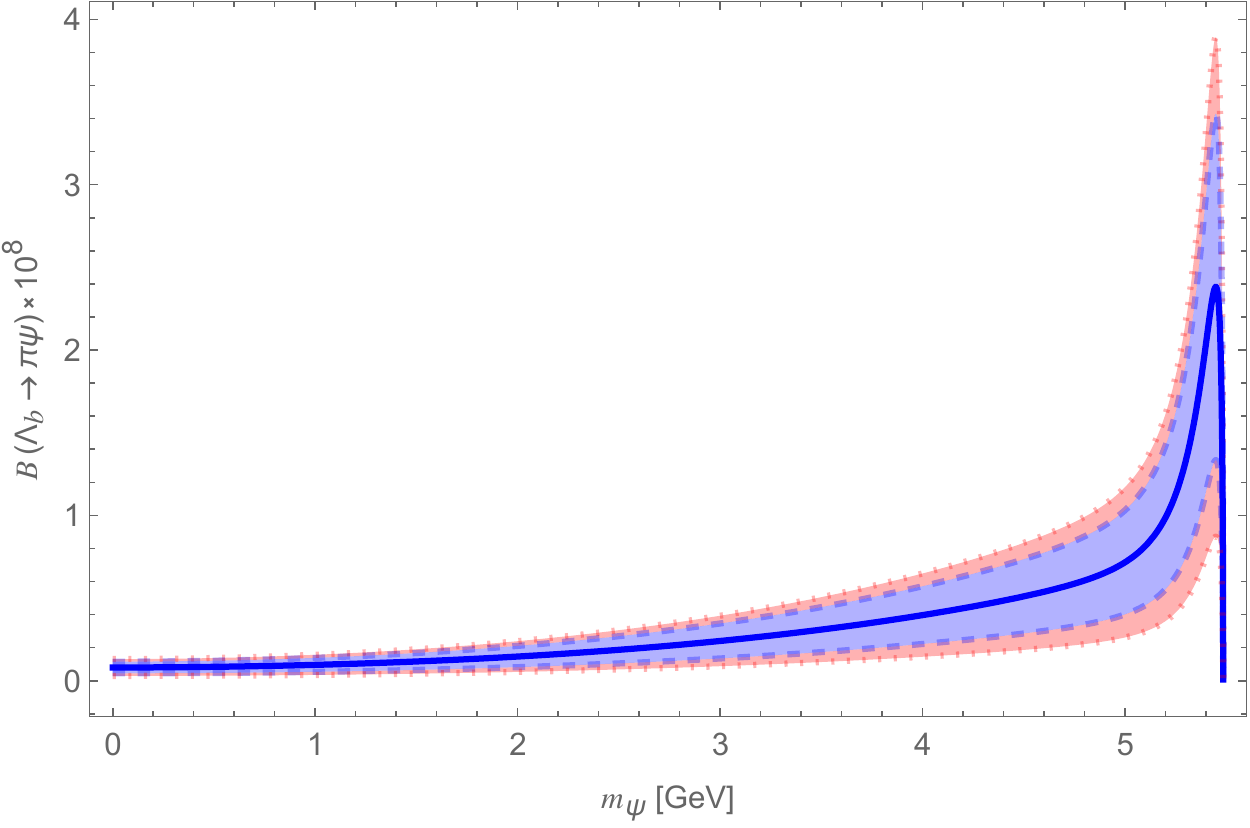}
\includegraphics[width=0.48\textwidth]{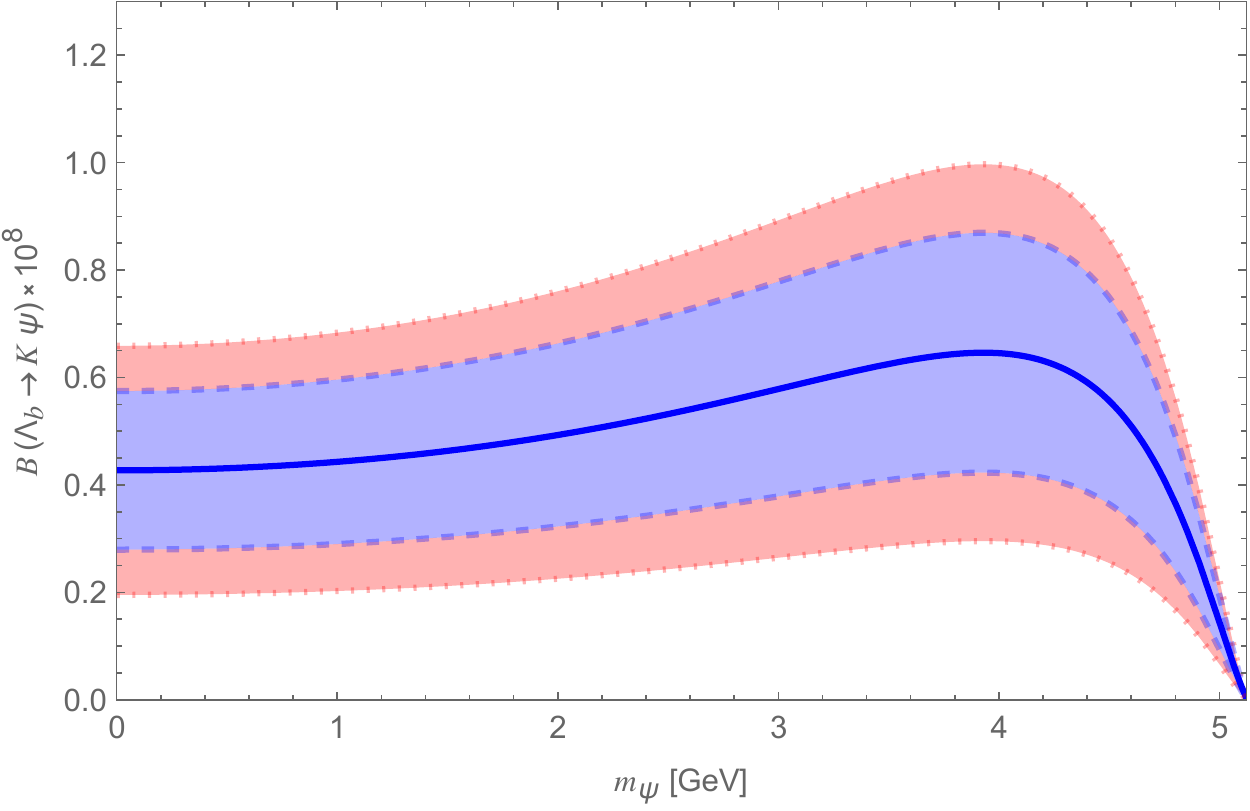}
\includegraphics[width=0.48\textwidth]{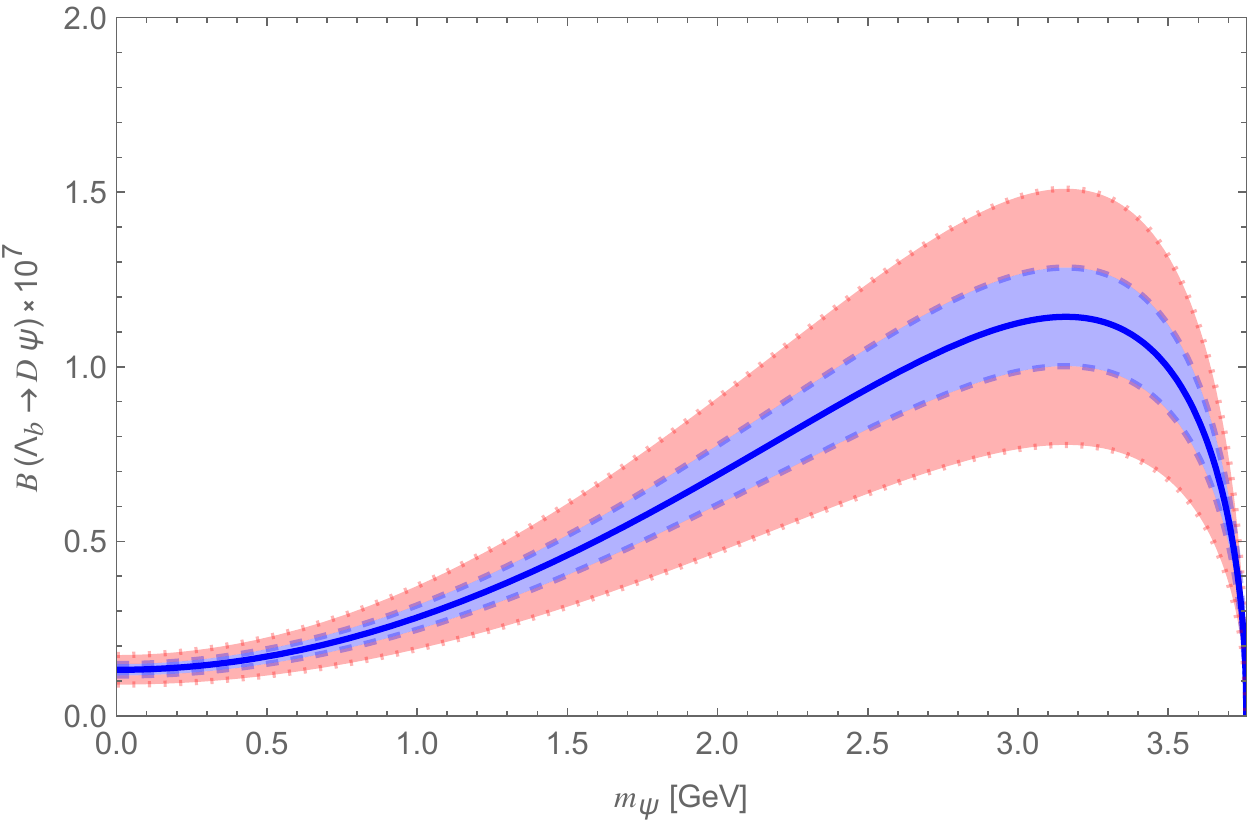}
\caption{The upper limits of branching fractions for $\Lambda_b \to M \psi$ as functions of $m_{\psi}$ in the type-I model. The blue and red bands denote the uncertainties from the Borel parameter and the $G_{(uq)}$ from Eq.(\ref{eq:Guqupper}).}
\label{fig:BrvsmpsitypeI}
\end{figure}

\begin{figure}
\centering
\includegraphics[width=0.48\textwidth]{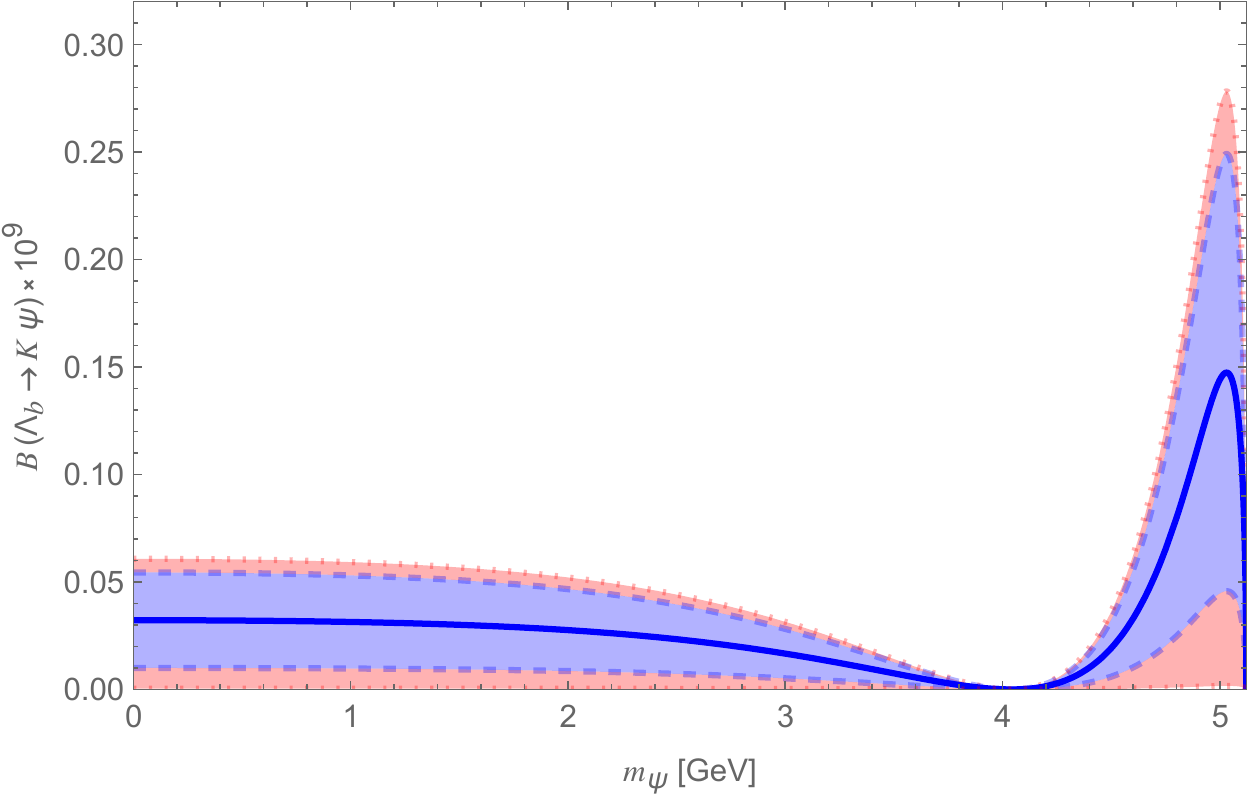}
\includegraphics[width=0.48\textwidth]{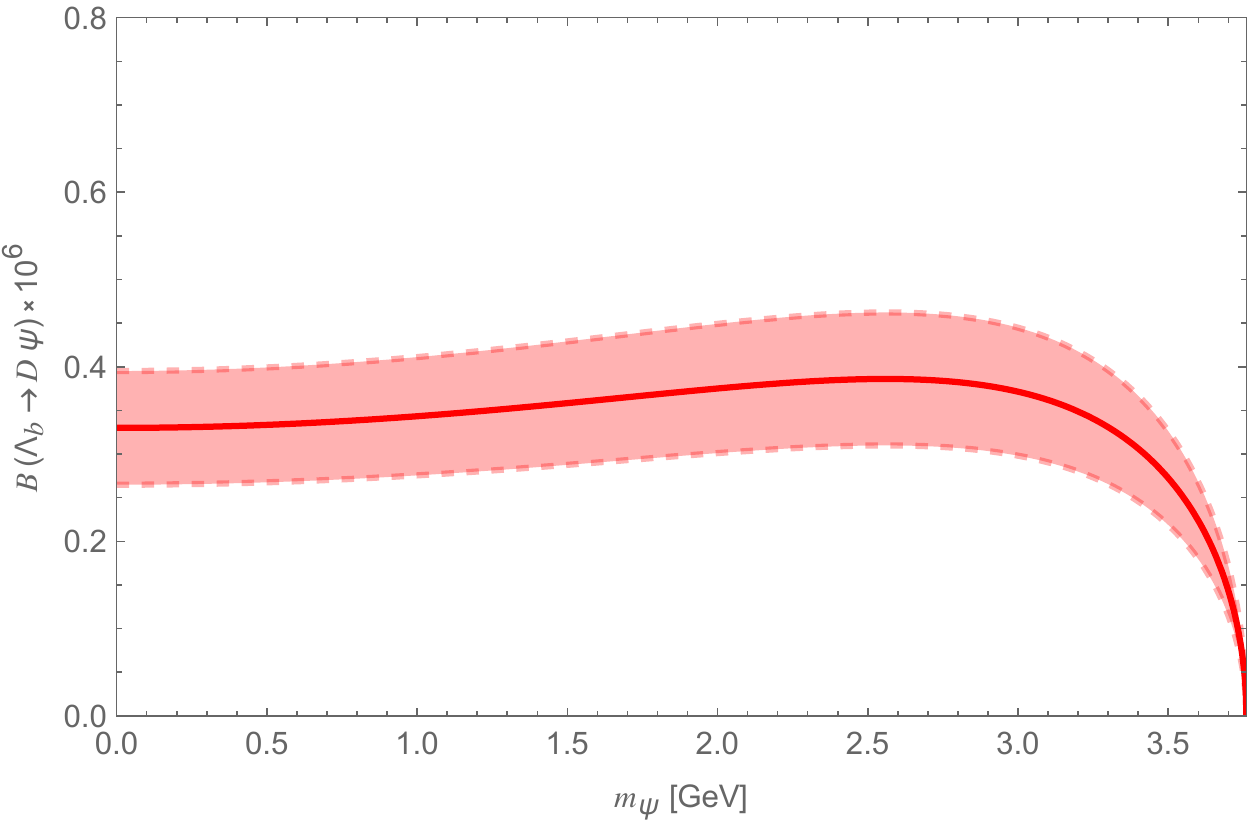}
\caption{The upper limits of branching fractions for $\Lambda_b \to M \psi$ as functions of $m_{\psi}$ in the type-II model. The blue and red bands denote the uncertainties from the Borel parameter and the $G_{(uq)}$ from Eq.(\ref{eq:Guqupper}).}
\label{fig:BrvsmpsitypeII}
\end{figure}

With the use of form factors $F_{1,2}$ obtained above, we can calculate the decay widths of $\Lambda_b \to M \psi$ decays. The decay width formula reads as
\begin{align}
&\Gamma[\Lambda_b \to M \psi]\nonumber\\
=& \frac{G_{(uq)}^2|\vec q|}{8 m_{\Lambda_b}^2(2\pi)^5}\left[(m_{\Lambda_b}^2+m_{\psi}^2-m_{M}^2) \left(F_1^2(m_{\psi}^2)+\frac{m_{\psi}^2}{m_{\Lambda_b}^2}F_2^2(m_{\psi}^2)\right)+4 m_{\psi}^2 F_1(m_{\psi}^2)F_2(m_{\psi}^2)\right],
\end{align}
which is a function of the dark baryon mass $m_{\psi}$. $|\vec q|$ is the 3-momentum magnitude of the dark baryon in the center of mass frame:
\begin{align}
|\vec q |=\frac{1}{2 m_{\Lambda_b}}\sqrt{(m_{\Lambda_b}^2-(m_{\psi}+m_M)^2)(m_{\Lambda_b}^2-(m_{\psi}-m_M)^2)}.
\end{align}
The upper limits of the unknown coupling constants $G_{(uq)}$ have been determined in Ref.\cite{Shi:2023riy} through a study of $B$ meson semi-inclusive decays into baryons and a dark baryon, which read as
\begin{align} 
{\rm Type \ I}:~~~~& G_{ud}^2<(1.8\pm0.35)\times 10^{-14} {\rm GeV}^{-4},~~~G_{us}^2<(3.75\pm0.74)\times 10^{-14} {\rm GeV}^{-4},\nonumber\\
& G_{cd}^2<(1.06\pm 0.21)\times 10^{-12} {\rm GeV}^{-4},~~~G_{cs}^2<(1.63\pm0.33)\times 10^{-12} {\rm GeV}^{-4};\nonumber\\
{\rm Type\ II}:~~~~& G_{us}^2<(1.07\pm 0.21) \times 10^{-11} {\rm GeV}^{-4},~~~ G_{cs}^2<(3.62\pm0.72)\times 10^{-10} {\rm GeV}^{-4}.\label{eq:Guqupper}
\end{align}
Here we take $G_{(uq)}$ as their upper limit values to calculate the decay width. Accordingly, the upper limits of branching fractions for $\Lambda_b \to M \psi$ as functions of $m_{\psi}$ are shown in Fig.\ref{fig:BrvsmpsitypeI} and Fig.\ref{fig:BrvsmpsitypeII}  in the type I and II models, respectively. The blue and red bands denote the uncertainties from the Borel parameter and the $G_{(uq)}$ from Eq.(\ref{eq:Guqupper}).

\begin{figure}
\centering
\includegraphics[width=1.0\textwidth]{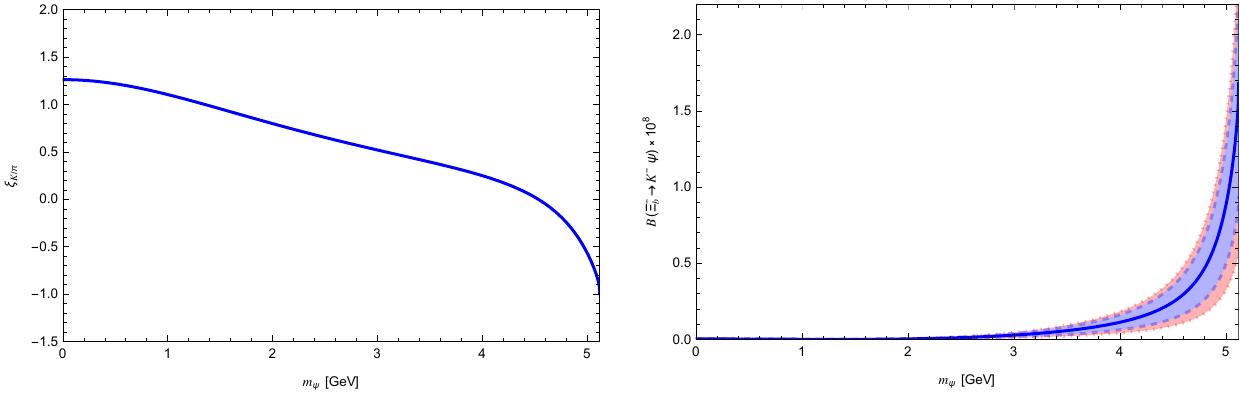}
\caption{The $\xi_{K/\pi}$ as a function of $m_{\psi}$ from Eq.(\ref{eq:xifromBr}), where the range of $m_{\psi}$ is chosen as $0<m_{\psi}<(m_{\Lambda_b}-m_K)$ (left). The branching fractions of $\Xi_b^- \to K^- \psi$ as functions of $m_{\psi}$ in the type-I model (right). The blue and red bands denote the uncertainties from the Borel parameter and the $G_{(ud)}$ from Eq.(\ref{eq:Guqupper}).}
\label{fig:xiKpi}
\end{figure}

In Sec.\ref{sec:SU3}, a SU(3) analysis is performed to predict the amplitudes of all the  anti-triplet bottom baryon decays into a meson and dark baryon. In the type-I model, using Eq.(\ref{eq:defofxi}) one can further determine $\xi_{K/\pi}$ by the ratio of branching fractions:
\begin{align}
\frac{1}{2}(1+\xi_{K/\pi})^2 \lambda_{s/d}^2=\frac{{\cal B}(\Lambda_b^0 \to K^0 \psi)}{{\cal B}(\Lambda_b^0 \to \pi^0 \psi)},\label{eq:xifromBr}
\end{align}
if the phase space volume of $\Lambda_b^0 \to K^0 \psi$ is assumed to be the same as that of $\Lambda_b^0 \to \pi^0 \psi$.  The $\xi_{K/\pi}$ as a function of $m_{\psi}$ is shown in the left diagram of Fig.\ref{fig:xiKpi}, where the range of $m_{\psi}$ is chosen as $0<m_{\psi}<(m_{\Lambda_b}-m_K)$ and $G_{ud}, G_{us}$ are taken as the center values in Eq.(\ref{eq:Guqupper}).
Now the branching fractions of all the decay channels listed in Table \ref{Tab:SU3amp} can be expressed in the units of  ${\cal B}(\Lambda_b^0 \to \pi^0 \psi)$ and ${\cal B}(\Lambda_b^0 \to K^0 \psi)$ for the type-I and II models respectively, which are shown in Table \ref{Tab:SU3ampBr}. Particularly, the branching fraction of the charged decay $\Xi_b^- \to K^- \psi$ in the type-I model is shown in the right diagram of Fig.\ref{fig:xiKpi}. The branching fraction of $\Xi_b^- \to \pi^- \psi$ has the same shape as $\Xi_b^- \to K^- \psi$ but a factor $\lambda_{s/d}^2$ should be timed on it.
\begin{table}
  \caption{The amplitudes of all the anti-triplet bottom baryon decays into a light meson and dark baryon are expressed in the units of  ${\cal B}(\Lambda_b^0 \to \pi^0 \psi)$ and ${\cal B}(\Lambda_b^0 \to K^0 \psi)$ for the type-I and II models, respectively.}
\label{Tab:SU3ampBr}
\begin{tabular}{|c|c|c|c|}
\hline
\hline
Channel & Type-I Br. & Channel & Type-I Br.  \tabularnewline 
\hline 
$\Lambda_b^0 \to \pi^0 \psi$ & 1 & $\Lambda_b^0 \to K^0 \psi$ & $\frac{1}{2}(1+\xi_{K/\pi})^2 \lambda_{s/d}^2$ \tabularnewline 
$\Xi_b^0 \to {\bar K}^0  \psi$ & $\frac{1}{2}(1+\xi_{K/\pi})^2$ & $\Xi_b^0 \to \pi^0  \psi$ & $\frac{1}{4}(1-\xi_{K/\pi})^2 \lambda_{s/d}^2$ \tabularnewline 
$\Xi_b^- \to K^-  \psi$ & $\frac{1}{2}(1-\xi_{K/\pi})^2$ & $\Xi_b^- \to \pi^-  \psi$ & $\frac{1}{2}(1-\xi_{K/\pi})^2 \lambda_{s/d}^2$ \tabularnewline 
$\Lambda_b^0 \to \eta  \psi$ & $\frac{1}{3}\xi_{K/\pi}^2$ & $\Xi_b^0 \to \eta  \psi$ & $\frac{1}{12}(1-3\xi_{K/\pi})^2 \lambda_{s/d}^2$ \tabularnewline 
\hline 
Channel & Type-II Br. & Channel & Type-II Br.  \tabularnewline 
\hline 
$\Lambda_b^0 \to \pi^0 \psi$ & 0 & $\Lambda_b^0 \to K^0 \psi$ & 1 \tabularnewline 
$\Xi_b^0 \to {\bar K}^0  \psi$ & $\lambda_{s/d}^{-2}$ & $\Xi_b^0 \to \pi^0  \psi$ & $1/2$ \tabularnewline 
$\Xi_b^- \to K^-  \psi$ & $\lambda_{s/d}^{-2}$ & $\Xi_b^- \to \pi^-  \psi$ & $1$ \tabularnewline 
$\Lambda_b^0 \to \eta  \psi$ & $\frac{2}{3}\lambda_{s/d}^{-2}$ & $\Xi_b^0 \to \eta  \psi$ & $1/6$ \tabularnewline 
\hline 
\end{tabular}
\end{table}
The branching fractions of the charged decays in the type-II model are simply proportional to ${\cal B}(\Lambda_b^0 \to K^0 \psi)$ so we do not plot them here.

\section{Conclusion}
\label{sec:conclusion}
In this work, we have studied the decays of Heavy baryon into a pseudoscalar meson and a dark baryon using the recently developed $B$-Mesogenesis scenario, where the two types of effective Lagrangians proposed by the scenario are both considered. The decay amplitudes of $\Lambda_b^0$ have been calculated by LCSR using its light-cone distribution amplitudes. The decay amplitudes of $\Xi_b^{0,\pm}$ has been related with those of $\Lambda_b^0$ through a flavor SU(3) analysis. In the numerical calculation, the uncertainties of threshold parameter and the Borel parameter are both considered. The values of effective coupling constants in the $B$-Mesogenesis are taken as their upper limits that obtained from our previous study on the inclusive decay. The upper limits of the decay branching fractions of $\Lambda_b^0, \Xi_b^{0,\pm} \to M \psi$ are presented as functions of the dark baryon mass, which will be tested by future experimental detections.

\section*{Acknowledgements}
The work of Y.J. Shi is supported by Natural Science Foundation of China under Grant No.12305103, and Opening Foundation of Shanghai Key Laboratory of Particle Physics and Cosmology under Grant No.22DZ2229013-2. The work of Y. Xing is supported by National Science Foundation of China under Grant No.12005294.  The work of Z.P. Xing is supported by China Postdoctoral Science Foundation under Grant No.2022M72210.

\begin{appendix}
	
\section{Analytical Results in the Type-I model}\label{app:analyI}

In this appendix, we present the analytical results of the quark-gluon level calculation for the correlation function defined in Eq.(\ref{eq:correfunc}). In the type-I model, the twist-3s contribution to the correlation function for the $\Lambda_b \to \pi$ decay is
\begin{align}
{\cal B}\{\Pi_{\Lambda_b\to \pi}^{QCD(3s)}\}(T,q)=&\frac{i}{4\sqrt{2}}f_{\Lambda_b}^{(1)}\int _0^{2 s_0^{\Lambda_b}}d\omega \int_0^1 du\ \theta(\Sigma-s_m)\theta(s_{\rm th}-\Sigma)e^{-\Sigma/T^2}\nonumber\\
&\times \psi_{3s}(\omega,u) P_R \left[m_{\Lambda_b}\omega \slashed q-\frac{m_{\Lambda_b}\omega}{m_{\Lambda_b}-u\omega}q^2\right]\nonumber\\
&+\frac{i}{4\sqrt{2}}f_{\Lambda_b}^{(1)}\int _0^{2 s_0^{\Lambda_b}}d\omega \int_0^1 du\ \theta({\bar \Sigma}-s_m)\theta(s_{\rm th}-{\bar \Sigma})e^{-{\bar \Sigma}/T^2}\nonumber\\
&\times \psi_{3s}(\omega,u) P_R \left[(\bar u \omega (2 q\cdot v)-2 p\cdot q -q^2)\frac{\omega m_{\Lambda_b}}{m_{\Lambda_b}-\bar u \omega}+m_{\Lambda_b} \omega \slashed q\right]u_{\Lambda_b}(v),
\end{align}
where $\Sigma=\Sigma_c|_{m_c = 0}$ and ${\bar \Sigma}=\Sigma|_{u\to \bar u}$. The twist-$3\sigma$ contributions are
\begin{align}
{\cal B}\{\Pi_{\Lambda_b\to \pi}^{QCD(3\sigma)}\}(p,q)
=&-\frac{i}{4\sqrt{2}}f_{\Lambda_b}^{(1)}\int _0^{2 s_0^{\Lambda_b}}d\omega \int_0^1 du\ \theta(\Sigma-s_m)\theta(s_{\rm th}-\Sigma)e^{-\Sigma/T^2}\nonumber\\
&\times \psi_{3s}(\omega,u) P_R \left[m_{\Lambda_b}\omega \slashed q-\frac{m_{\Lambda_b}\omega}{m_{\Lambda_b}-u\omega}q^2\right]u_{\Lambda_b}(v)\nonumber\\
&-\frac{i}{4\sqrt{2}}f_{\Lambda_b}^{(1)}\int _0^{2 s_0^{\Lambda_b}}d\omega \int_0^1 du\ \theta({\bar \Sigma}-s_m)\theta(s_{\rm th}-{\bar \Sigma})e^{-{\bar \Sigma}/T^2}\nonumber\\
&\times \psi_{3\sigma}(\omega,u) P_R \left[(\bar u \omega (2 q\cdot v)-2 p\cdot q -q^2)\frac{\omega m_{\Lambda_b}}{m_{\Lambda_b}-\bar u \omega}+m_{\Lambda_b} \omega \slashed q\right]u_{\Lambda_b}(v)\nonumber\\
&-\frac{i}{2\sqrt{2}}f_{\Lambda_b}^{(1)}\frac{\partial}{\partial M^2}\int _0^{2 s_0^{\Lambda_b}}d\omega \int_0^1 du\ {\bar\psi}_{3\sigma}(\omega,u) e^{-(\Sigma+M^2)/T^2}\nonumber\\
&\times \theta((\Sigma+M^2)-s_m)\theta(s_{\rm th}-(\Sigma+M^2))\left(\frac{m_{\Lambda_b}}{m_{\Lambda_b}-u\omega}\right)^2\nonumber\\
&\times P_R [A(p^2,q^2)+B(p^2,q^2)\slashed q]u_{\Lambda_b}(v){\Big |}_{M^2=0,\  p^2=\Sigma+M^2}\nonumber\\
&+\frac{i}{\sqrt{2}}f_{\Lambda_b}^{(1)}\frac{\partial}{\partial M^2}\int _0^{2 s_0^{\Lambda_b}}d\omega \int_0^1 du\ {\bar\psi}_{3\sigma}(\omega,u) e^{-({\bar \Sigma}+M^2)/T^2}\nonumber\\
&\times \theta(({\bar \Sigma}+M^2)-s_m)\theta(s_{\rm th}-({\bar \Sigma}+M^2))\left(\frac{m_{\Lambda_b}}{m_{\Lambda_b}-\bar u\omega}\right)^2\nonumber\\
&\times (p\cdot q-\bar u \omega v\cdot q)P_R (m_{\Lambda_b}-\bar u \omega -\slashed q)u_{\Lambda_b}(v){\Big |}_{M^2=0,\  p^2=\bar \Sigma+M^2}\  ,
\end{align}
The twist-3s and twist-$3\sigma$ contributions  to the correlation function for the $\Lambda_b \to K$ decay are
\begin{align}
{\cal B}\{\Pi_{\Lambda_b\to K}^{QCD(3s)}\}(T,q)=&-\frac{i}{4}f_{\Lambda_b}^{(1)}\int _0^{2 s_0^{\Lambda_b}}d\omega \int_0^1 du\ \theta({\bar \Sigma}-s_m)\theta(s_{\rm th}-{\bar \Sigma})e^{-{\bar \Sigma}/T^2}\nonumber\\
&\times \psi_{3s}(\omega,u) P_R \left[(\bar u \omega (2 q\cdot v)-2 p\cdot q -q^2)\frac{\omega m_{\Lambda_b}}{m_{\Lambda_b}-\bar u \omega}+m_{\Lambda_b} \omega \slashed q\right]u_{\Lambda_b}(v),\nonumber\\
{\cal B}\{\Pi_{\Lambda_b\to K}^{QCD(3\sigma)}\}(p,q)
=&\frac{i}{4}f_{\Lambda_b}^{(1)}\int _0^{2 s_0^{\Lambda_b}}d\omega \int_0^1 du\ \theta({\bar \Sigma}-s_m)\theta(s_{\rm th}-{\bar \Sigma})e^{-{\bar \Sigma}/T^2}\nonumber\\
&\times \psi_{3\sigma}(\omega,u) P_R \left[(\bar u \omega (2 q\cdot v)-2 p\cdot q -q^2)\frac{\omega m_{\Lambda_b}}{m_{\Lambda_b}-\bar u \omega}+m_{\Lambda_b} \omega \slashed q\right]u_{\Lambda_b}(v)\nonumber\\
&-i f_{\Lambda_b}^{(1)}\frac{\partial}{\partial M^2}\int _0^{2 s_0^{\Lambda_b}}d\omega \int_0^1 du\ {\bar\psi}_{3\sigma}(\omega,u) e^{-({\bar \Sigma}+M^2)/T^2}\nonumber\\
&\times \theta(({\bar \Sigma}+M^2)-s_m)\theta(s_{\rm th}-({\bar \Sigma}+M^2))\left(\frac{m_{\Lambda_b}}{m_{\Lambda_b}-\bar u\omega}\right)^2\nonumber\\
&\times (p\cdot q-\bar u \omega v\cdot q)P_R (m_{\Lambda_b}-\bar u \omega -\slashed q)u_{\Lambda_b}(v){\Big |}_{M^2=0,\  p^2=\bar \Sigma+M^2}.
\end{align}
The twist-2, 4 contributions  to the correlation function for the $\Lambda_b \to \pi, K$ decays vanishes in the chiral limit $m_u=m_d=0$.

\section{Analytical Results in the Type-II model}\label{app:analyII}
In the type-II model, the $\Lambda_b \to \pi$ decay is forbidden in the flavor SU(3) limit. The twist-3s and twist-$3\sigma$ contributions  to the correlation function for the $\Lambda_b \to K$ decay are
\begin{align}
{\cal B}\{\Pi_{\Lambda_b\to K}^{QCD(3s)}\}(T,q)=&-\frac{i}{2}f_{\Lambda_b}^{(1)}\int _0^{2 s_0^{\Lambda_b}}d\omega \int_0^1 du\ \theta({\bar \Sigma}-s_m)\theta(s_{\rm th}-{\bar \Sigma})e^{-{\bar \Sigma}/T^2}\nonumber\\
&\times \psi_{3s}(\omega,u) \frac{\omega m_{\Lambda_b}}{m_{\Lambda_b}-\bar u \omega} P_R [\bar u \omega v\cdot q -p\cdot q]u_{\Lambda_b}(v),\nonumber\\
{\cal B}\{\Pi_{\Lambda_b\to K}^{QCD(3\sigma)}\}(p,q)
=&\frac{i}{2}f_{\Lambda_b}^{(1)}\int _0^{2 s_0^{\Lambda_b}}d\omega \int_0^1 du\ \theta({\bar \Sigma}-s_m)\theta(s_{\rm th}-{\bar \Sigma})e^{-{\bar \Sigma}/T^2}\nonumber\\
&\times \psi_{3\sigma}(\omega,u) \frac{\omega m_{\Lambda_b}}{m_{\Lambda_b}-\bar u \omega} P_R [\bar u \omega v\cdot q -p\cdot q]u_{\Lambda_b}(v)\nonumber\\
&-2 i f_{\Lambda_b}^{(1)}\frac{\partial}{\partial M^2}\int _0^{2 s_0^{\Lambda_b}}d\omega \int_0^1 du\ {\bar\psi}_{3\sigma}(\omega,u) e^{-({\bar \Sigma}+M^2)/T^2}\nonumber\\
&\times \theta(({\bar \Sigma}+M^2)-s_m)\theta(s_{\rm th}-({\bar \Sigma}+M^2))\left(\frac{m_{\Lambda_b}}{m_{\Lambda_b}-\bar u\omega}\right)^2\nonumber\\
&\times (v\cdot p-\bar u \omega)(p\cdot q -\bar u \omega v\cdot q) P_R u_{\Lambda_b}(v){\Big |}_{M^2=0,\  p^2=\bar \Sigma+M^2}.
\end{align}
The twist-2+4 contributions  to the correlation function for the $\Lambda_b \to D$ decay are
\begin{align}
{\cal B}\{\Pi_{\Lambda_b\to D}^{QCD(2+4)}\}(p,q)
=&\frac{i}{2}m_c f_{\Lambda_b}^{(2)}\int _0^{2 s_0^{\Lambda_b}}d\omega \int_0^1 du\ \theta(\Sigma_c-s_m)\theta(s_{\rm th}-\Sigma_c)e^{-\Sigma_c/T^2}\nonumber\\
&\times \psi_{2}(\omega,u) \frac{\omega m_{\Lambda_b}}{m_{\Lambda_b}-\bar u \omega}(v\cdot q)P_R u_{\Lambda_b}(v)\nonumber\\
&+\frac{i}{2}m_c f_{\Lambda_b}^{(2)}\frac{\partial}{\partial M^2}\int _0^{2 s_0^{\Lambda_b}}d\omega \int_0^1 du\ ({\bar\psi}_{2}-{\bar\psi}_{4})(\omega,u) e^{-({\Sigma_c}+M^2)/T^2}\nonumber\\
&\times \theta(({\Sigma_c}+M^2)-s_m)\theta(s_{\rm th}-({\Sigma_c}+M^2))\left(\frac{m_{\Lambda_b}}{m_{\Lambda_b}-\bar u\omega}\right)^2\nonumber\\
&\times (p\cdot q-u\omega v\cdot q) P_R u_{\Lambda_b}(v){\Big |}_{M^2=0,\  p^2=\Sigma_c+M^2}.
\end{align}
The twist-3 contributions  to the correlation function for the $\Lambda_b \to D$ decay are
\begin{align}
{\cal B}\{\Pi_{\Lambda_b\to D}^{QCD(3s)}\}(p,q)
=&-i f_{\Lambda_b}^{(1)}\int _0^{2 s_0^{\Lambda_b}}d\omega \int_0^1 du\ \theta(\Sigma_c-s_m)\theta(s_{\rm th}-\Sigma_c)e^{-\Sigma_c/T^2}\nonumber\\
&\times (\psi_{3s}+\psi_{3\sigma})(\omega,u) \frac{\omega m_{\Lambda_b}}{m_{\Lambda_b}-\bar u \omega}(u\omega v\cdot q-p\cdot q)P_R u_{\Lambda_b}(v)\nonumber\\
&+2 i f_{\Lambda_b}^{(1)}\frac{\partial}{\partial M^2}\int _0^{2 s_0^{\Lambda_b}}d\omega \int_0^1 du\ {\bar\psi}_{3\sigma}(\omega,u) e^{-({\Sigma_c}+M^2)/T^2}\nonumber\\
&\times \theta(({\Sigma_c}+M^2)-s_m)\theta(s_{\rm th}-({\Sigma_c}+M^2))\left(\frac{m_{\Lambda_b}}{m_{\Lambda_b}-\bar u\omega}\right)^2\nonumber\\
&\times (v\cdot p-u\omega) (p\cdot q-u\omega v\cdot q) P_R u_{\Lambda_b}(v){\Big |}_{M^2=0,\  p^2=\Sigma_c+M^2}.
\end{align}

\end{appendix}

\end{document}